\documentclass{PoS}
\usepackage{url}
\usepackage{amsmath}
\usepackage{floatrow}
\usepackage{caption}
\usepackage{siunitx}
\usepackage{ucs}
\usepackage[utf8x]{inputenc}
\usepackage{sidecap}
\DeclareCaptionLabelSeparator*{colonnewline}{:\\}

\title{Astrophysical sources and acceleration mechanisms}

\ShortTitle{Astrophysical sources}

\author{Martina Adamo,$^a$ Silvia Pietroni,$^{bc}$ and \speaker{Maurizio Spurio}\thanks{Talk based on M. Spurio's lecture experiences at the University of Bologna on \textit{Astroparticle Physics} and on the book \textit{M. Spurio, Probes of Multi-messenger Astrophysics}, part of the book series \textit{Astronomy and Astrophysics Library (AAL)}. \cite{Spurio:2018knn}} $^{de}$\\
    \\
    \llap{$^a$}Departamento de Física, Universidad de Burgos, Facultad de Ciencias, Plaza Misael Ba\~nuelos s.n. 09001-Burgos, Spain\\
    \llap{$^b$}Dipartimento di Fisica “E.R. Caianiello”, Università di Salerno, Via Giovanni Paolo II 132, I-84084 Fisciano, Italy\\
    \llap{$^c$}Istituto Nazionale di Fisica Nucleare, Sezione di Napoli, Via Cintia, 80126, Napoli, Italy\\
    \llap{$^d$}Dipartimento di Fisica e Astronomia dell'Universit\`a di Bologna, Viale Berti Pichat 6/2, 40127 Bologna, Italy\\
    \llap{$^e$}INFN - Sezione di Bologna, Viale Berti-Pichat 6/2, 40127 Bologna, Italy

 \\ 
    \\
    E-mail: \email{madamo@ubu.es}, \email{marti.adamo@gmail.com},\\
    \hspace{33pt}\email{spietroni@unisa.it},\\
    \hspace{34pt}\email{maurizio.spurio@unibo.it}}

\abstract{\vspace{30pt} Multi-messenger astronomy provides for the observation of the same astronomical event with different kind of telescopes at the same time: optical observations, X-rays, $\gamma$-rays, neutrinos and, most recently, gravitational waves are just few examples of the several points of view from which an astronomical event can be observed and analyzed. Cosmic rays play an important role in multi-messenger astronomy and, for this reason, it is important to deepen the study of their sources and to understand the mechanisms behind their acceleration in astronomical environments.}

\FullConference{Corfu Summer Institute 2021 "School and Workshops on Elementary Particle Physics and Gravity" (CORFU2021)\\
		29 August - 9 October 2021\\
		Corfu, Greece}

\begin{document}

\section{Introduction}

With the beginning of the use of optical spectroscopy in astronomy in the 19\textsuperscript{th} century, the observations of light from stars allowed a deeper understanding of the Universe, since until then it was based principally on electromagnetic observations. Over the decades, observational techniques have been improved, enabling to expand the range of detected wavelengths more and more. Despite this, the Earth's atmosphere prevents a huge part of the electromagnetic radiation to reach the Earth. Therefore, there was another step forward with X-rays observations in the 60s-70s, thanks to the lunch of X-ray detectors via rockets outside the atmosphere. The combined observation of the electromagnetic radiation form X-rays to radio-waves using different detectors is called \textit{multi-wavelength astronomy}.

Afterwards, the advancement of technology allowed to investigate the Universe from different perspectives, i.e., neutrinos, cosmic rays, high-energy photons and, the most recent, gravitational waves, marking the beginning of the era of the \textit{multi-messenger astronomy}. This approach provides for the observation of the same astronomical event with different kind of telescopes at the same time: one of the most stunning example is the observation of the merging of two spiraling neutron stars in August 2017. This event was first detected by the LIGO-VIRGO Collaboration through gravitational waves and independently by the $\gamma$-ray Burst Monitor on the Fermi Satellite and by the INTEGRAL Satellite through $\gamma$-rays. Right after this combined observations, all the other telescopes (such as optical and neutrinos telescopes) focused on the research of the source of these signals, which was located in the galaxy NGC 4993, about 40 Mpc from the Earth. The event was then studied in the following weeks through its electromagnetic radiation, which allowed to know the composition of the elements synthesized during the coalescence \cite{LIGOneutron, 2017ApJ...850L..35A}.

Cosmic rays are an important protagonist of the multi-messenger astronomy: they are very-high-energy particles, mostly protons and fully-ionized atomic nuclei (up to iron nuclei) plus a small percentage of solitary electrons, photons, high-energy neutrinos ($>$TeV), positrons and antiprotons. Usually, we refer to "cosmic rays" only to protons and nucleii \cite{courvoisier2012high, gassier}. Cosmic rays are accelerated in various astrophysical processes and therefore can be classified with respect to their source:

\begin{itemize}
    \item cosmic rays originate from the Sun, called \textit{solar energetic particles};
    \item cosmic rays originate from our galaxy, called \textit{galactic cosmic rays};
    \item cosmic rays originate from external galaxies, called \textit{extragalactic cosmic rays}.
\end{itemize}

All these particles travel across the space and reach the Earth passing through the atmosphere. The collisions between the particles of cosmic rays and atoms and molecules of Earth's atmosphere (mainly oxygen and nitrogen) produce a cascade of lighter particles (called \textit{air shower}), such as pions, muons, electrons and neutrinos. \cite{sokolsky2020introduction}

Cosmic rays are also classified into \textit{primary} and \textit{secondary}, depending on whether their detection is made respectively before or after the interaction with the interstellar medium or the Earth’s atmosphere. As the flux of cosmic rays strongly decreases with the increasing of the energy, different measurement techniques are needed \cite{gassier}. These techniques are described in Sec. \ref{primasezione}, with the illustration of some cosmic rays experiments (PAMELA, KASCADE, PAO). In Sec. \ref{spectra} we introduce the physical concepts of integral and differential fluxes from a mathematical point of view, focusing on the energy spectrum of cosmic rays. In Sec. \ref{terzasezione} and Sec \ref{quartasezione} we describe the cosmic ray diffusion in the Milky Way and the main acceleration mechanisms (magnetic mirrors, first- and second-order Fermi acceleration models). In Sec. \ref{quintasezione} we focus more on the supernova remnants and cosmic ray sources, discussing the point of view of their energy. Sec. \ref{sestasezione} is about the extragalactic sources of cosmic rays and their acceleration. In Sec. \ref{settimasezione} we describe secondary particles more in detail, $\gamma$-rays and neutrinos inter alia: their detection and the (possible) correlation with other events. Conclusions are in Sec. \ref{conclusioni}.

\section{Direct and indirect cosmic rays detection}\label{primasezione}

Since primary cosmic rays produce secondary particles after the interaction with the atmosphere, it is possible to detect them only at high altitude using balloon-borne instruments or satellites in space, before they produce secondary particles. These methods permit a direct study of primary cosmic rays and for this reason they are called methods of \textit{direct detection} \cite{grounddetector} and they allow us to study the chemical composition, i.e., the relative fraction of different nuclei present in cosmic radiation and sometimes even the isotopic composition.

Other important quantities whose measurements direct detectors deal with are energy and momentum: for this goal, detectors are equipped with \textit{calorimeters} and \textit{spectrometers}, with some constraints in weight and size as they need to be mounted on balloons and satellites \cite{poggiani2016high}.

\begin{itemize}
    \item Calorimeters are apparatus that measure the energy of the particles. A calorimeter collects and measures the energy of an entering particle as it interacts with the instrument ad generates a particle shower of secondary particles whose energy is dissipated via excitation/ionization of the absorbing material of which the calorimeter is composed.
    
    \item Magnetic spectrometers are devices that measure the momentum of the particles, through their rigidity, i.e., their resistance to deflection by external magnetic fields generated by a solenoid. The rigidity can be measured up to a maximum value which depends on the magnetic field strength and on the precision with which the particle path through the detector is measured.
\end{itemize}

At energies above $\sim$ 10$\div$100 TeV, it is no longer possible to use direct experiments to detect primary cosmic rays: these detectors are characterized by a small geometrical factor, hence if the cosmic ray flux is below few tens of particles per $\mathrm{m}^2$ per year (such is for high energy cosmic rays, i.e., above $10^{15}$ eV) direct detection becomes nearly impossible. For this reason direct experiments are replaced by \textit{indirect detection} via ground-based experiments, which measure secondary particles by distributing detectors over an instrumented surface covering up to several thousands of $\mathrm{km}^2$ and for long periods of time. As we said in the previous section, the interaction of cosmic rays particles with Earth's atmosphere produces a cascade of particles called air showers, so these experiments are called \textit{extensive air shower (EAS) arrays}. \cite{baloon, griender, meas, direct}

\subsection{The PAMELA experiment}

As an example of direct experiments, we consider the PAMELA apparatus. \textit{PAMELA} is the acronym of "Payload for Antimatter Matter Exploration and Light-nuclei Astrophysics". It was the first satellite-based experiment for the direct detection of cosmic rays: PAMELA was a module of the Russian Resurs-DK1 satellite which was lunched on June 2006 through a Soyuz-U rocket and was operational until 2016.

\begin{figure}[!ht]
\centering
\includegraphics[width=0.7\textwidth]{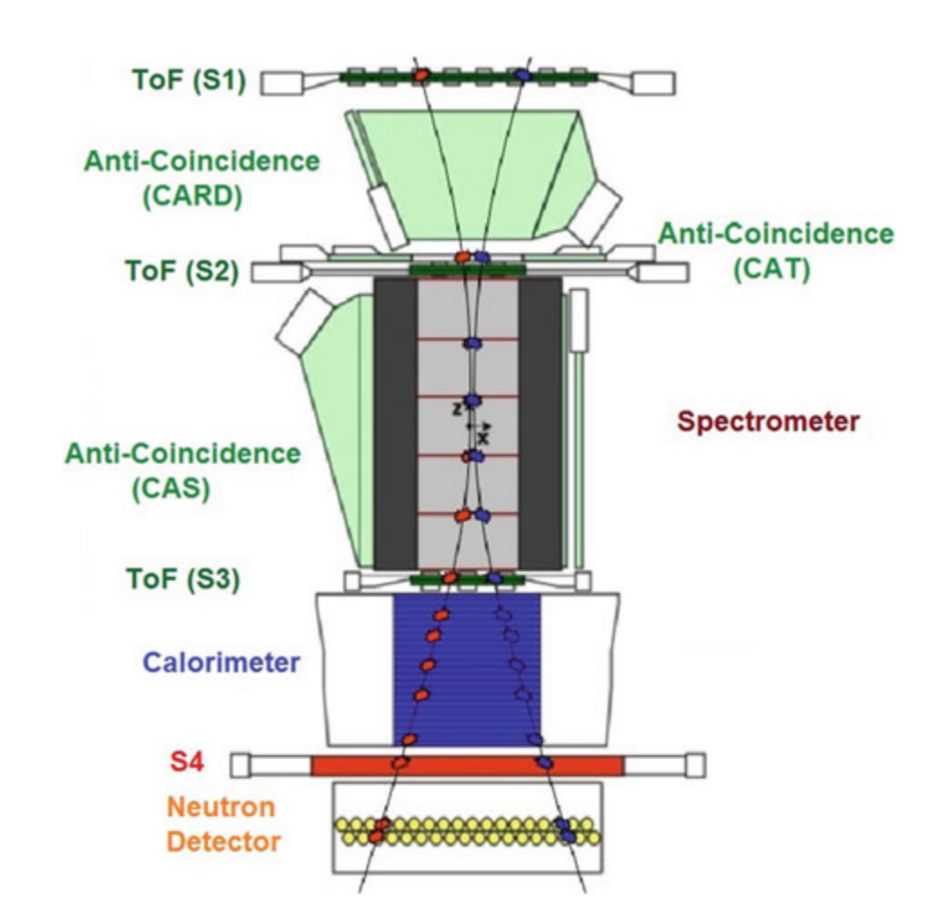} 
\setcaptionwidth{0.8\textwidth}
\caption{\footnotesize{Cross-section of the PAMELA detector with two opposite-sign charged particles. Combined measurements from the magnetic spectrometer, calorimeter, time of flight system, and neutron detectors shown here distinguish the incident particles by their charge, momentum, and mass. \cite{pcoll}}}
\label{pamela}
\end{figure}

This cosmic rays detector was composed of different subdetectors (see Fig. \ref{pamela}): the central part was a magnetic spectrometer, whose magnetic field was generated by a $0.43$ T permanent magnet and a tracking system for the measurement of the particles paths whose resolution was of $3$ $\mathrm{\mu}$m, resulting in a maximum detectable rigidity of $\sim 1$ TV; below the spectrometer there was a calorimeter, composed of an alternation of silicon and tungsten planes, with a total depth of 16.3 radiation lengths and 0.6 interaction lengths. The calorimeter was supported by a plastic scintillator system for the identification of high-energy electrons and a neutron detection system in order to improve the electromagnetic/hadronic discrimination capabilities of the calorimeter. \cite{anom, pam, pam2, pam3, pam10}

\subsection{The KASCADE experiment}

As an example of indirect experiments, we consider \textit{KASCADE} (KArlsruhe Shower Core and Array DEtector) experiment. It was an European experiment located in Germany which started in 1996, afterwards extended by the KASCADE-Grande, and ended in 2013. It was an extensive air shower experiment array to study the cosmic ray primary composition and the hadronic interactions in the energy range $10^{14}\div 10^{17}$ eV. The main goal was to detect secondary particles as electrons, muons at four energy thresholds, and energy and number of hadrons. This was accomplished with several detectors: the leading ones were a field array (for electrons and muons detection) organized in an array of 252 detector stations, each of which formed by shielded and unshielded detectors disposed in a square grid of $200\times 200$ $\mathrm{m}^2$, and a muon-tracking detector formed by 4 plastic scintillators, each of which covering a surface of $0.8$ $\mathrm{m}^2$. In 2003 KASCADE-Grande became operational, and the measurement of primary cosmic rays was extended up to
$10^{18}$ eV. \cite{kask1, kask2, kask3, kask4}

\begin{figure}[!ht]
\centering
\includegraphics[width=0.7\textwidth]{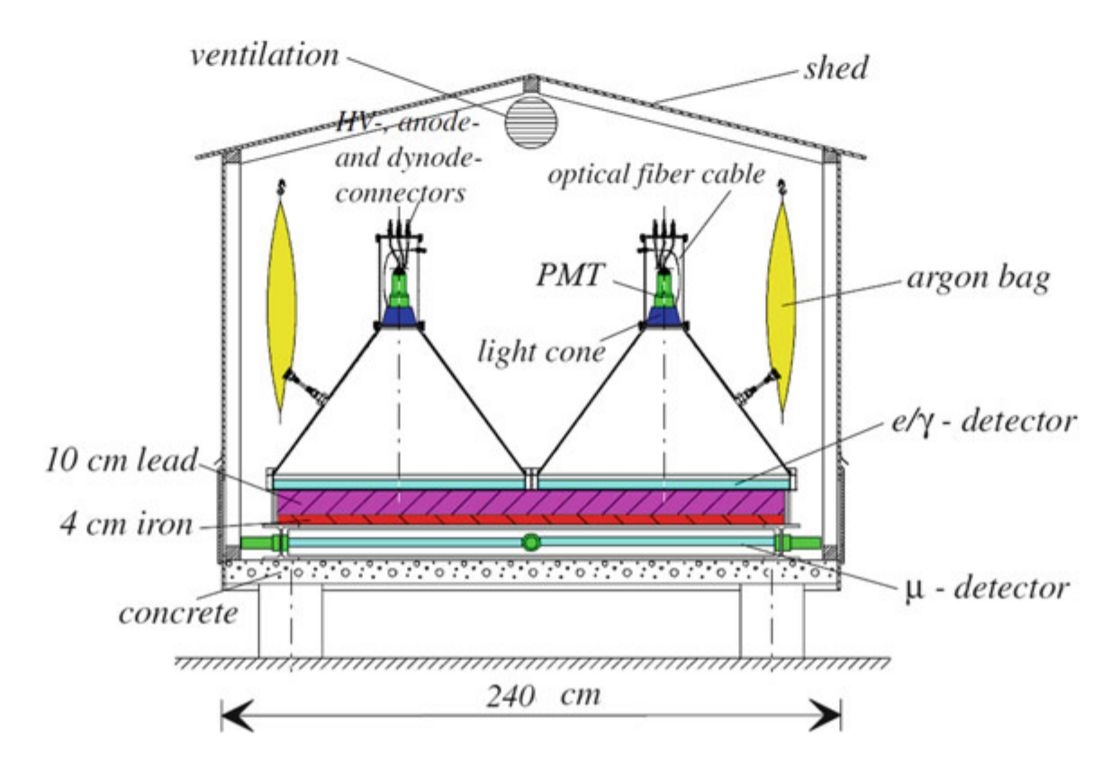} 
\setcaptionwidth{0.8\textwidth}
\caption{\footnotesize{Schematic view of one of the 256 sample array detector stations of the KASCADE experiment. \cite{kask}}}
\label{kascade}
\end{figure}

\subsection{The Pierre Auger Observatory}

The \textit{Pierre Auger Observatory (PAO)} is the largest cosmic rays observatory ever built: it is located in Argentina, completed in 2008 (but collecting data since 2004) and still operational. It detects ultra-high-energy cosmic rays (i.e., energies beyond $10^{18}$ eV) reaching the Earth using both fluorescence and surface array detection techniques and this allows very precise measurements.

\begin{figure}[!ht]
\centering
\includegraphics[width=0.7\textwidth]{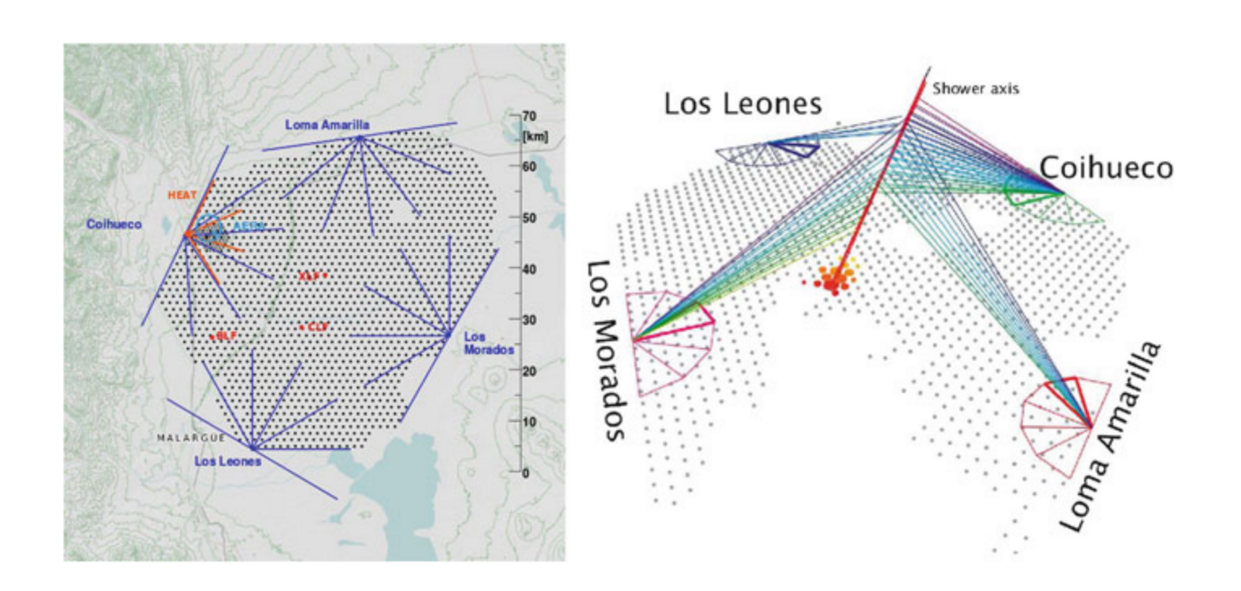} 
\setcaptionwidth{0.8\textwidth}
\caption{\footnotesize{The Pierre Auger Observatory. On the left, the position and field-of-view of the fluorescence detector eyes surrounding the array is displayed. The dots mark the positions of the 1600 surface detector tanks. On the right an event detected by the four fluorescence detector eyes in coincidence with the surface detector array. \cite{pao}}}
\label{pao}
\end{figure}

Fluorescence detectors are specialised telescopes which detect the fluorescing effect due to incoming high-energy primary cosmic rays: air showers generated by every high-energy sub-atomic particle passing through the Earth's atmosphere are made up of longitudinally spread particles whose speed is close to the speed of light. These planes of high-speed particles create the fluorescing effect interacting with the atmosphere. The fluorescence detectors system of PAO is also equipped with three tiltable fluorescence telescopes positioned at high altitudes, called High Elevation Auger Telescopes (HEAT), which permit to detect also fluorescence due to primary cosmic rays with lower energy (down to $10^{17}$ eV).

The surface detector array of the PAO is formed by 1600 water Cherenkov detectors disposed on an area of 3000 $\mathrm{km}^2$, plus 60 other detectors disposed in a denser array: the latter is coupled with the HEAT fluorescence detectors. These arrays are sensitive to electrons, positrons, muons and $\gamma$-rays. \cite{2017arXiv170806592T, 2017Sci...357.1266P, 2017ApJ...850L..35A}

\subsection{Abundances of elements in the solar system and in cosmic rays}

\begin{figure}[!ht]
\centering
\includegraphics[width=0.7\textwidth]{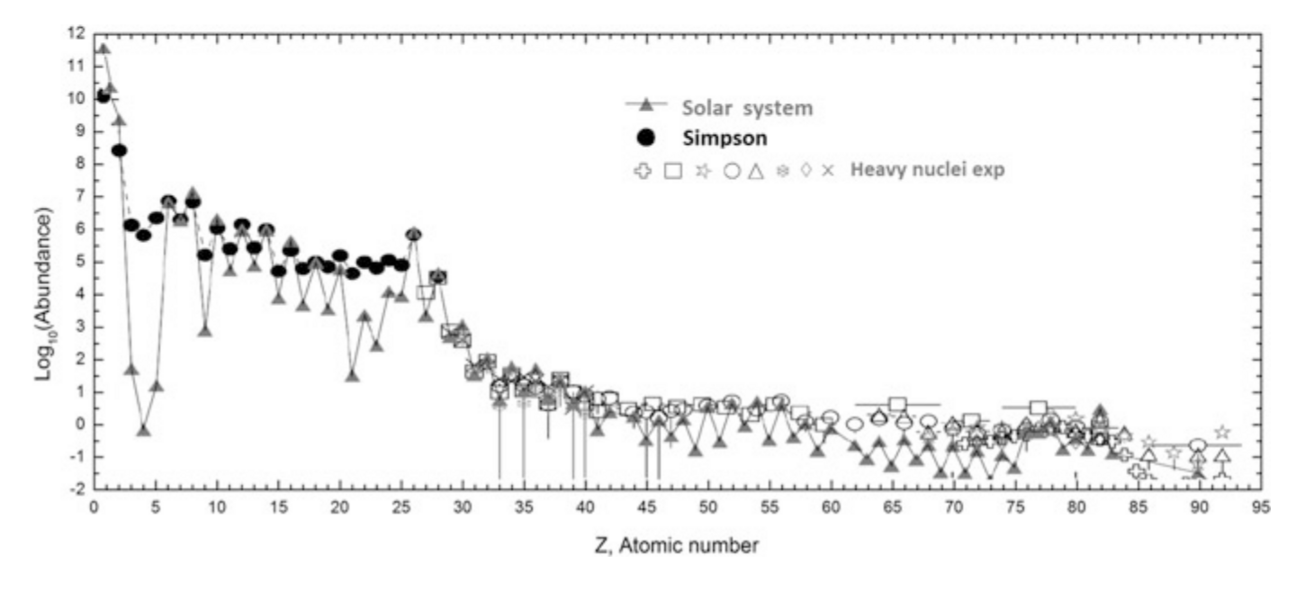} 
\setcaptionwidth{0.8\textwidth}
\caption{\footnotesize{Relative abundance of nuclei in cosmic rays as a function of their atomic number $Z$ at energies around 1 GeV $\mathrm{n}^{-1}$, normalized to Si=100. Abundances for nuclei with $Z \leq 28$ are drawn according to Simpson (1983). \cite{simp} Heavier nuclei are measured by different experiments, as reported in Bl\"umer et al. (2009). \cite{blum} The abundance of elements (triangles) in the solar system according to Lodders (2003) is also shown. \cite{abu4}}}
\label{comparison}
\end{figure}

It is interesting to compare the relative abundances of elements in cosmic rays as a function of the atomic number $Z$ with the abundances of the same elements in our solar system, because it can give many information about the origin and the history of the particles that form cosmic rays. This comparison is shown in Fig. \ref{comparison}: in the plot we notice the alternation of abundances of adjacent atomic numbers, as we expect from energy levels of adjacent nuclei in the nuclear binding energy curve; moreover, even if with some differences, data from cosmic ray and solar system abundances are very similar, and their differences are within $20\%$ in most cases. This means that it is reasonable to assume that cosmic rays sources have the same chemical composition of our solar system and that the mechanism behind the formation of cosmic rays is similar to the one that originated the solar system. \cite{abu, abu2, abu3, abu4, abu5}

\section{The cosmic rays spectrum}\label{spectra}

\subsection{Differential and integral fluxes}

A typical measurement of cosmic ray telescopes allows to determine the number of incident particles $N$ per unit of time on the detector surface $A$ at a given solid angle $d\Omega$: in general the area seen by incoming particles is a function of the arrival direction, which can be uniquely identified by two angles $\theta, \phi$ (zenith and azimuth angles), i.e., $A=A(\theta,\phi)=A(\Omega)$. We can define the \textit{geometric factor} as 

\begin{equation}
    A\Omega\equiv\int\! d\Omega\ \ A(\Omega) \, .
\end{equation}

The detectors that measure also the energy of the particles permit to determine another quantity which is important in order to study the cosmic rays: that is the \textit{differential intensity of particles of a given energy in a given solid angle} or \textit{differential flux}, i.e., the number of particles $N$ per unit of area and time $A\cdot T$, in a given energy interval $dE$ and a given solid angle $d\Omega$

\begin{equation}
    \Phi (E)=\frac{d^2(\tfrac{N}{A\cdot T})}{d\Omega \ dE} \, .
\end{equation}

The differential flux can also be integrated over energy, from a threshold energy $E_0$ up to infinity

\begin{equation}
    \Phi (E>E_0)=\int_{E_0}^{+\infty}\!\! dE \ \frac{d^2(\tfrac{N}{A\cdot T})}{d\Omega \ dE}  =\frac{d(\tfrac{N}{A\cdot T})}{d\Omega} \, ,
\end{equation}

\noindent which is the \textit{integral intensity of particles with energy above $E_0$}, also called \textit{integral flux}. Since the arrival direction of cosmic rays is approximately isotropic, we can also integrate the differential flux over solid angle with respect to different surfaces, e.g., 

\begin{equation}
    \begin{aligned}
        \mathcal{F}_{sphere} & = \int_{sphere}\! \! d\Omega \ \ \Phi (E)= 4\pi \ \Phi (E) \, ,\\
        \mathcal{F}_{plane} & = \int_{plane} \!\! d\Omega \ \ \Phi (E)= \pi \ \Phi (E) \, .
    \end{aligned}
\end{equation}

\noindent These quantities can be integrated further over energy, from a threshold energy $E_0$ up to infinity, resulting in $\mathcal{F}(E>E_0)=\alpha \  \Phi(E>E_0)$ (where $\alpha$ is a constant which depends on the integration surface). \cite{jack}

\subsection{Energy spectra}

\begin{figure}[!ht]
\centering
\includegraphics[width=0.8\textwidth]{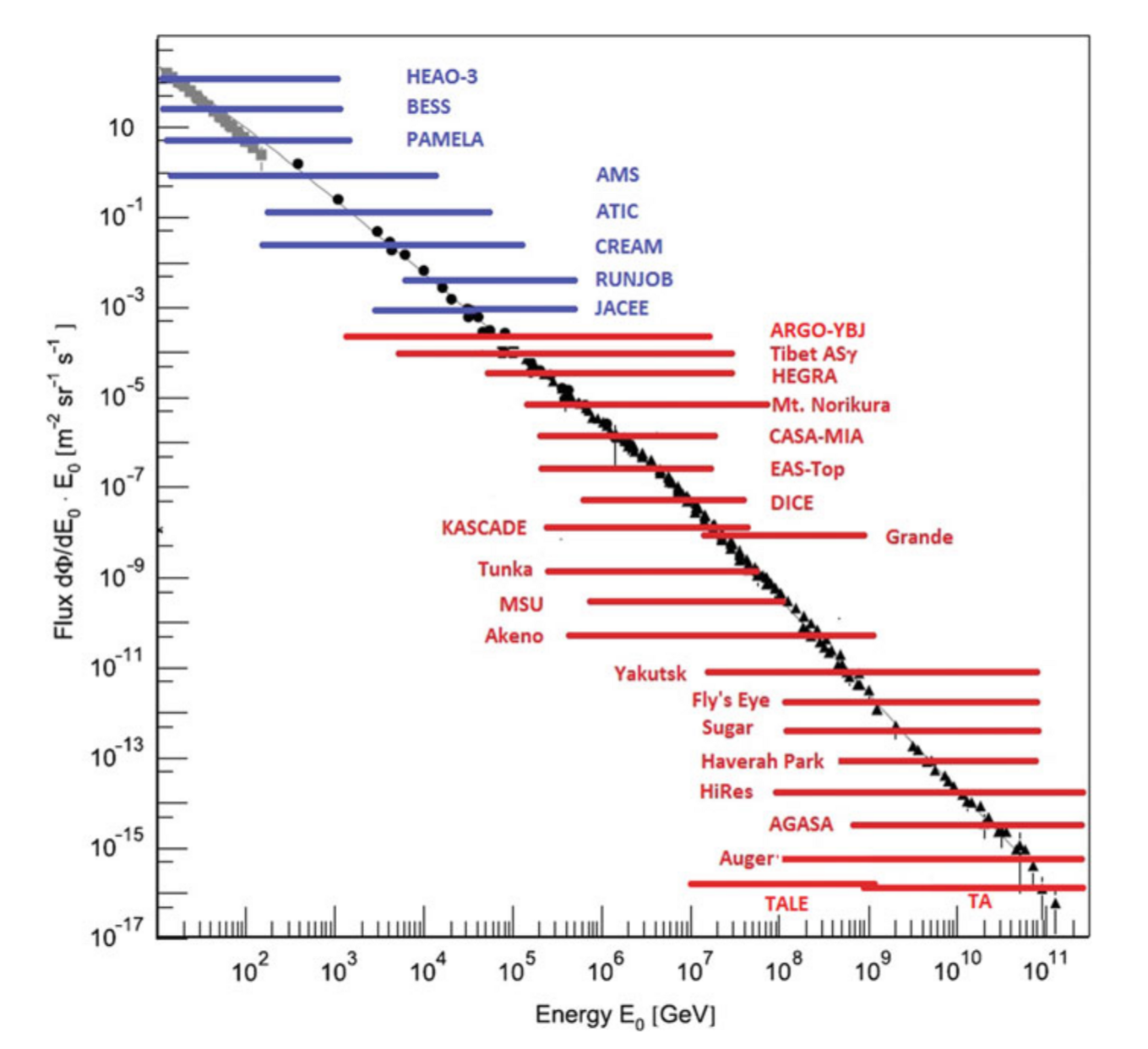} 
\setcaptionwidth{0.8\textwidth}
\caption{\footnotesize{Flux of cosmic rays as a function of energy. Below a few GeV, the contribution of particles coming from the Sun is not negligible. The energy range of the cosmic ray flux measured by some direct experiments is shown as a blue line and that measured by some indirect experiments as a red line.}}
\label{flux1}
\end{figure}

Each cosmic ray detector is able to measure the differential and the integral fluxes over a given energy interval: if we consider all of them, both direct and indirect experiments, we can cover several orders of magnitude in energy. The analytic interpolation of this data is called \textit{integral energy spectrum} (Fig. \ref{flux1}). As we can see, the flux decreases rapidly as the energy increases: this means that we measure many particles of low energy and very few particles of high energy. In particular we can identify three thresholds in the integral flux

\begin{align}
    \mathcal{F}(E>10^9\  \mathrm{eV}) & \simeq 1000 \  \mathrm{particles/s\  m}^2 \, , \nonumber \\ 
    \mathcal{F}(E>10^{15}\  \mathrm{eV}) & \simeq 1\  \mathrm{particle/year\  m}^2 \, , \\
    \mathcal{F}(E>10^{20}\  \mathrm{eV}) & \simeq 1\  \mathrm{particle/century\ km}^2\, . \nonumber
\end{align}

\noindent The first threshold concerns particles detected via direct experiments (primary cosmic rays), while the other two concern particles detected via indirect experiments (air showers).

\begin{figure}[!ht]
\centering
\includegraphics[width=0.8\textwidth]{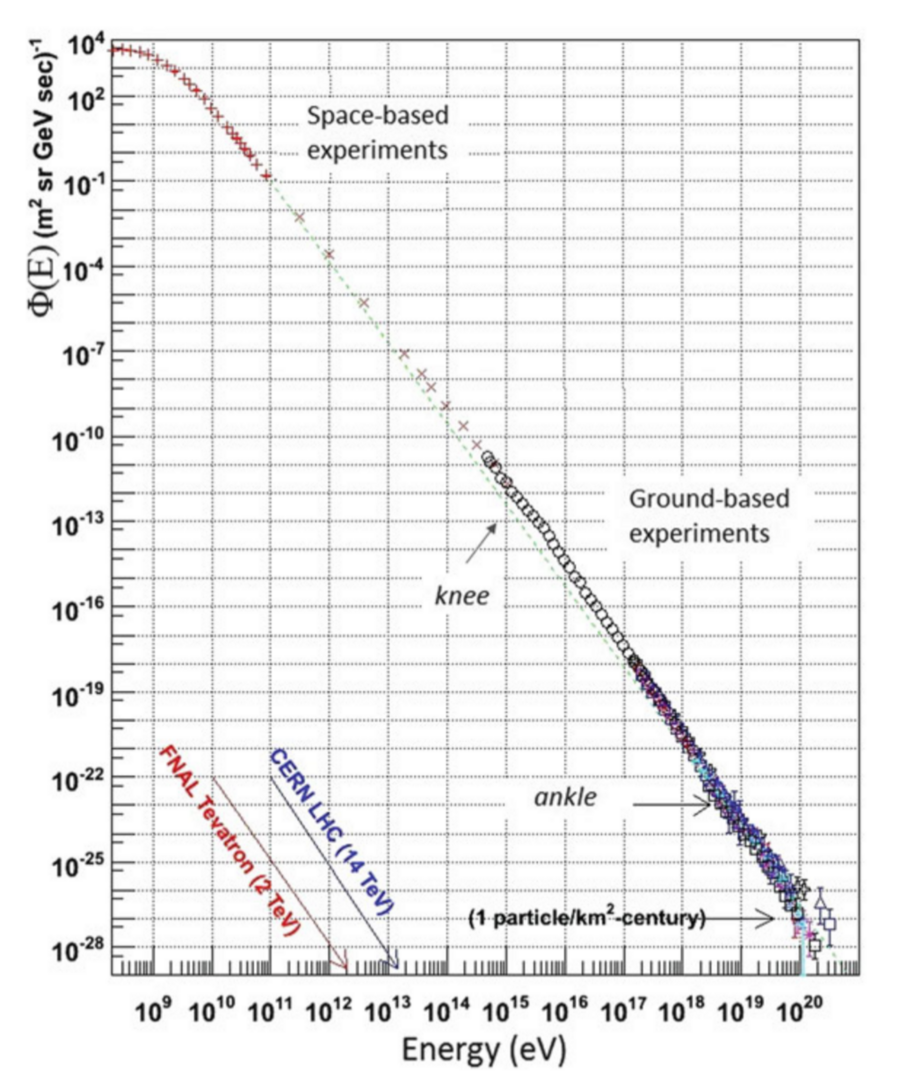} 
\setcaptionwidth{0.8\textwidth}
\caption{\footnotesize{The differential energy spectrum $\Phi (E)$ (units: particles/$\mathrm{m}^2$ sr s GeV) of cosmic rays over eleven decades of energy. The red/blue arrows indicate the equivalent center of mass energy reached at the Tevatron collider at Fermilab and at the LHC collider at CERN. Note that the spectrum is remarkably continuous over the whole energy interval, and that the flux on the y-axis covers 33 decades. The dashed line shows a $\mathrm{E}^{−3}$ spectrum.}}
\label{flux2}
\end{figure}

Similarly, from the information regarding the differential flux obtained in all the cosmic ray experiments, it is possible to reconstruct its distribution over the whole energy interval, which is called \textit{differential energy spectrum} (Fig. \ref{flux2}). Here we can identify two transitions point, where the slope of the spectrum changes, which are called the \textit{knee} and the \textit{ankle} of the spectrum. The knee corresponds to an energy of $\sim 3 \cdot 10^6$ GeV: below this point (lower energies) the flux decreases by a factor 50 when the energy increases by any order of magnitude, above this point (higher energies) the flux decreases by a factor 100 when the energy increases by a factor 10 up to the ankle. The ankle corresponds to an energy of $\sim 10^{10}$ GeV: above this point the spectra becomes flatter and it is thought that cosmic rays in this region have an extragalactic origin. Above $10^{20}$ eV very few cosmic rays have been observed.

\begin{figure}[!ht]
\centering
\includegraphics[width=0.9\textwidth]{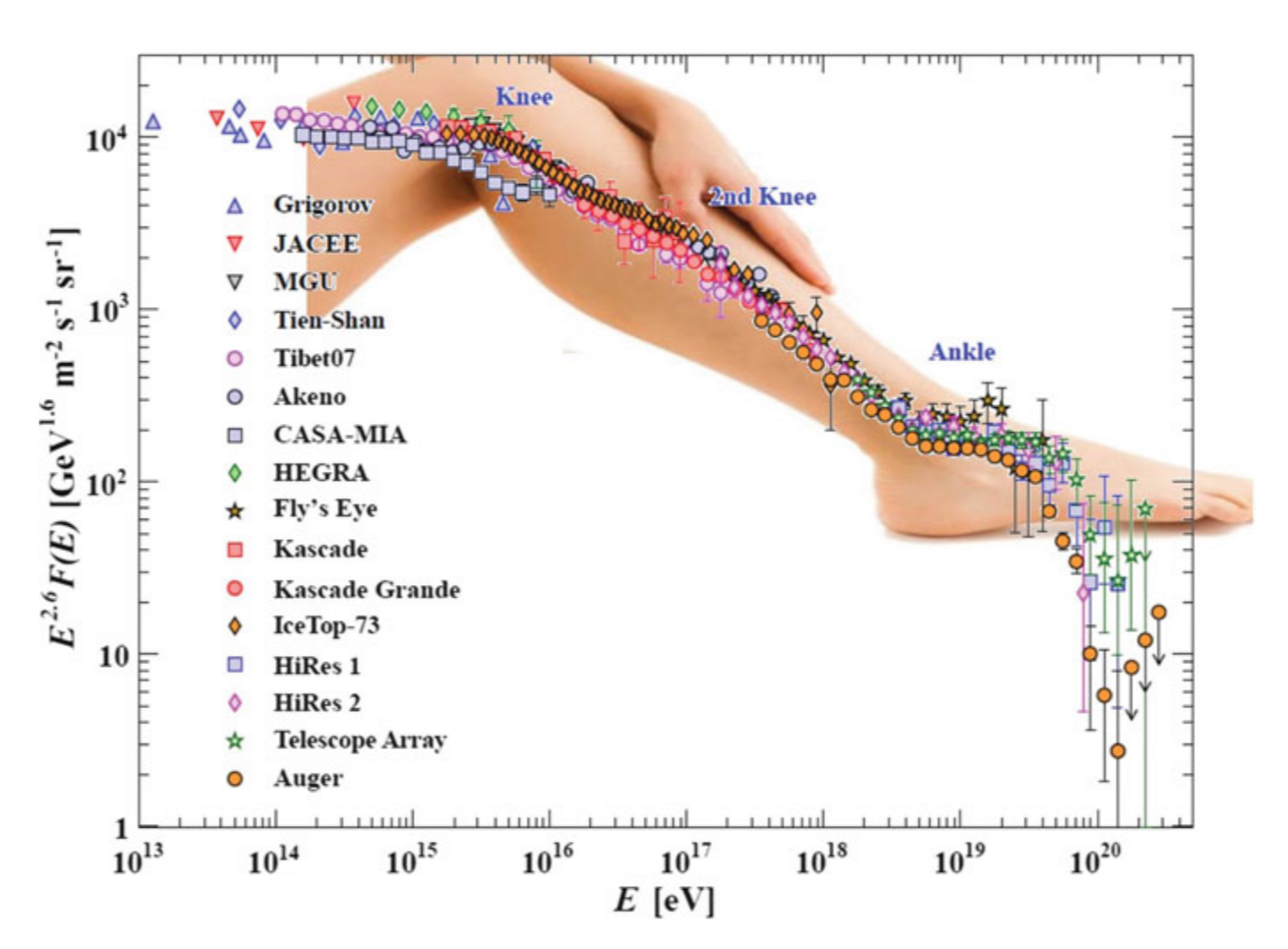} 
\setcaptionwidth{0.8\textwidth}
\caption{\footnotesize{The differential cosmic ray flux $\Phi (E)$ as measured by many direct and indirect cosmic ray experiments over eight decades of energy, almost the same as Fig. \ref{flux2}. Here, the flux is multiplied by a power of the energy: $\Phi (E) \cdot E^{2.6}$. The structures of the knee and ankle are more evident. Adapted from a figure from Sect. 27: Cosmic Rays of Beringer et al. (2012). \cite{review}}}
\label{flux3}
\end{figure}

At low energies (below the knee) the greatest contribution to the flux comes from cosmic rays originating from the Sun, while above few GeV this contribution becomes negligible and the energy spectrum can be approximated by a power low

\begin{equation}
    \frac{d\Phi}{dE}= A \ E^{-\alpha} \, ,
\end{equation}

\noindent where the parameter $\alpha\simeq 2.7$ is the slope of the cosmic ray spectrum, also called the \textit{differential spectral index of the cosmic ray flux}, and $A$ is a normalization factor. In a double-logarithmic representation, power laws are straight lines

\begin{equation}
    \log\left(\frac{d\Phi}{dE}\right)=\log A - \alpha \ \log E\, ,
\end{equation}

\noindent and $\alpha$ and $A$ becomes the slope of the line and the intercept with the y-axis respectively.

The ankle and knee are more evident if we multiply the y-axis by $E^{\beta}$, where $\beta$ is a real number, and in this case, using a double-logarithmic representation, straight lines have different slopes

\begin{equation}
    \log\left(\frac{d\Phi}{dE} \ E^\beta\right)=\log A - (\alpha-\beta)\log E\, .
\end{equation}

\noindent For example in Fig. \ref{flux3} it is shown the integral flux using $\beta=2.6$. \cite{wiebel, knee, review}

\section{Cosmic rays diffusion in the Milky Way} \label{terzasezione}

Milky Way is the galaxy in which our solar system is located: it is a spiral galaxy whose radius is $\sim 15$ kpc, approximately flat (the thickness is $\sim 200\div 300$ pc) but with a spheroidal structure in the center which contains a very massive black hole, i.e., $2\times 10^6\  \mathrm{M}_\odot$. The disk is composed for the most part by dust and gas (which are the cause of the absorption of the interstellar radiation) and by young stars. All the gases and the dust in the Galaxy are called \textit{interstellar matter}, which represents only the $5\div 10\%$ of the Galaxy mass, with an average density of $\sim 1\ \mathrm{proton}/\mathrm{cm}^3$. The Galaxy is also filled with magnetic fields, whose average intensity is $4\ \mu$G.

Galactic cosmic rays originate in \textit{source regions} and diffuse in the galactic magnetic field randomly, which is the cause of their large isotropy: the observed spectrum depends on the acceleration of the particles in the astrophysical sources and their propagation through the Galaxy. \cite{gal1, gal2, gal3}

\subsection{Motion of charged particles in a magnetic field}\label{magnetic field}

A particle with mass $m$ and charge $q$ that moves in a magnetic field $\vec{B}$ with a velocity $\vec{v}$, is affected by the Lorentz force

\begin{equation}
    m\ \Gamma\ \frac{d\vec{v}}{dt}=\frac{q}{c}\ \vec{v}\times \vec{B}\, , 
\end{equation}

\noindent where $\Gamma=\tfrac{E}{mc^2}$ is the Lorentz factor.

\begin{figure}[!ht]
\centering
\includegraphics[width=0.6\textwidth]{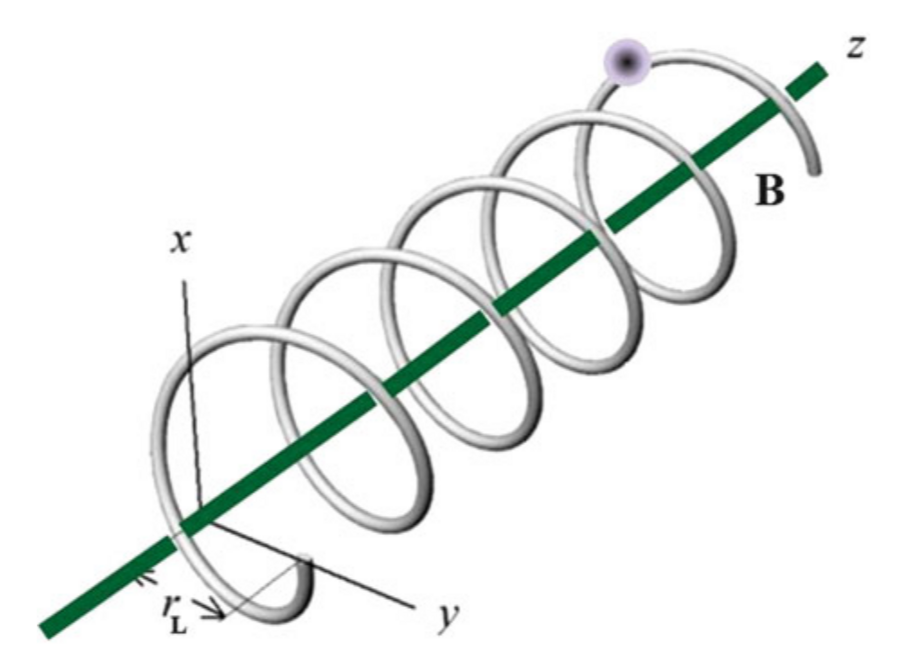} 
\setcaptionwidth{0.8\textwidth}
\caption{\footnotesize{The motion of a charged particle along the 
$\vec{B}$ field line.}}
\label{magnetic}
\end{figure}

\noindent If the field is static and uniform, the orbit is helicoidal in the direction of the field (Fig. \ref{magnetic}) and the radius of the orbit is called
\textit{Larmor radius}

\begin{equation} \label{larmor}
    r_L=\frac{v}{\omega_{\,L}}=\frac{\Gamma\,  m\, v\, c}{Z\, e\, B}=\frac{p\, c}{Z\, e\, B}\simeq\frac{E}{Z\, e\, B}\, ,
\end{equation}

\noindent where $\omega_{\,L}=\tfrac{q\,B}{\Gamma\, m\, c}$ is the angular frequency of the circular motion and $q=Z\,e$ is the charge of the particle ($e$ is the electric charge of the proton and $Z$ the atomic number); the last equality holds only for relativistic particles.

\begin{figure}[!ht]
\centering
\includegraphics[width=0.95\textwidth]{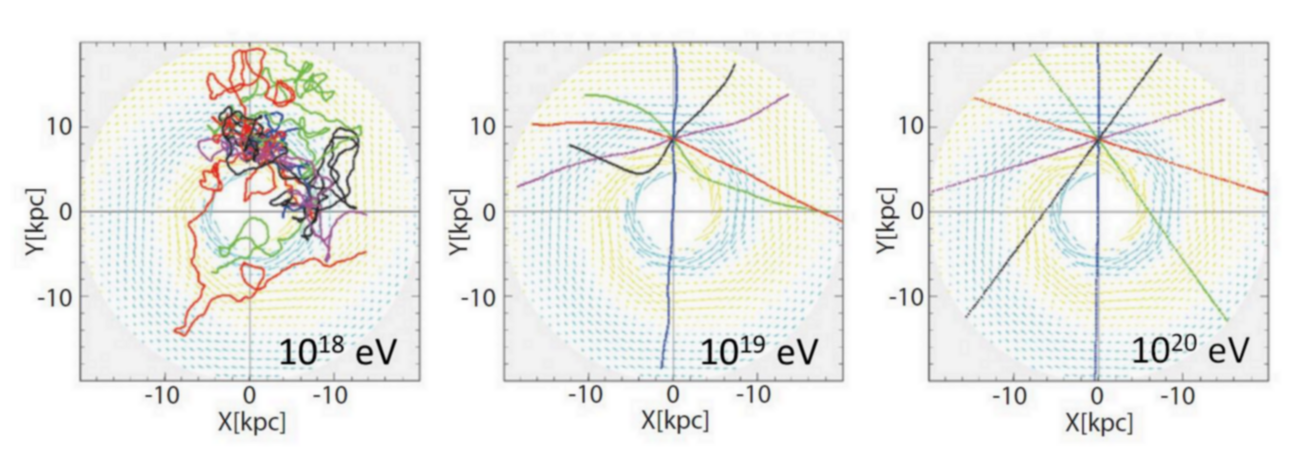} 
\setcaptionwidth{0.8\textwidth}
\caption{\footnotesize{Simulated trajectory of cosmic rays in the galactic magnetic field. Low-energy charged particles are bent by magnetic fields, but those above $10^{20}$ eV travel along almost straight trajectories. Credit: Prof. T. Ebisuzaki \cite{riken}}}
\label{galactic}
\end{figure}

For a typical galactic magnetic field $B\simeq 4 \ \mu$G and for a proton $Z=1$, therefore the Larmor radius depends only on the energy of the particle $E$, e.g.,

\begin{align}
    r_L(E=10^{12}\ \mathrm{eV}) & \simeq 10^{15} \ \mathrm{cm}= 3\cdot 10^{-4} \ \mathrm{pc} \, , \nonumber \\
    r_L(E=10^{15}\ \mathrm{eV}) & \simeq 10^{18} \ \mathrm{cm}= 0.3 \ \mathrm{pc} \, , \\
    r_L(E=10^{18}\ \mathrm{eV}) & \simeq 10^{21} \ \mathrm{cm}= 300 \ \mathrm{pc} \, , \nonumber
\end{align}

\noindent and this shows that the curvature radius is much smaller then the height of the galactic disc, at energies below $10^{18}$ eV: this is called \textit{galactic confinement}. For this reason, ultra-high energy cosmic rays are not of galactic origin (Fig. \ref{galactic}).

Cosmic rays are produced and accelerated in the source regions, which are characterized by large matter density and magnetic fields: when accelerated, high-energy electrons lose energy due to inverse Compton scattering, bremsstrahlung and synchrotron radiation. This is the reason why high-energy electrons are rare in cosmic rays. \cite{jack, spurio}

\section{Galactic accelerators and acceleration mechanisms} \label{quartasezione}

From experimental observations we know that cosmic rays have energies which typically are higher than the energies reachable with only blackbody radiations and thermal bremsstrahlung (thermal energies): this means that acceleration processes are needed to explain the data. Artificial accelerators use electric and magnetic fields, while in astrophysical environments, the matter in the state of plasma prevents static electric fields. Galactic cosmic rays, which have energies below the knee in the spectrum, are accelerated by very violent processes, the supernovae explosions; the acceleration of the bulk of cosmic rays is due to recursive stochastic mechanisms, which provide energy to low-energy particles with a large number of interactions between these particles and shock waves. This acceleration model, called \textit{standard model of cosmic ray acceleration}, describes very well the flux below the knee, but it fails in the description of the flux above and it requires additional models. \cite{sho1, sho2}

\subsection{Magnetic mirrors}

\begin{figure}[!ht]
\centering
\includegraphics[width=0.5\textwidth]{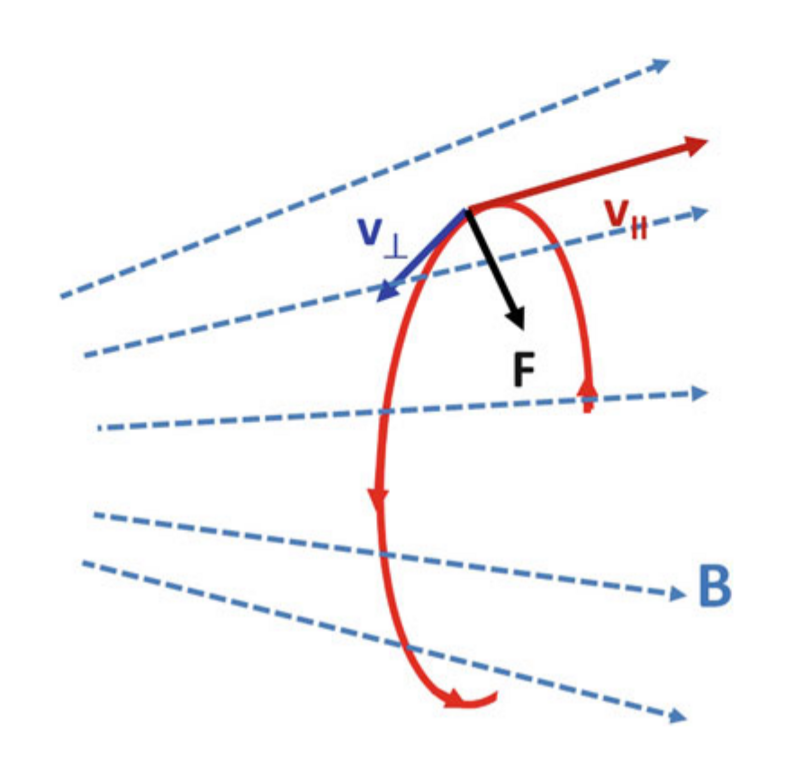} 
\setcaptionwidth{0.8\textwidth}
\caption{\footnotesize{Motion of a charged particle in a nonuniform magnetic field.}}
\label{mirror1}
\end{figure}

In Subsec. \ref{magnetic field} we have seen the behaviour of a charged particle moving in a uniform magnetic field: if the magnetic field is inhomogeneous, the motion will not be helicoidal anymore, since the irregularities of the field (called \textit{magnetic mirrors}) cause scatterings of the particle (Fig. \ref{mirror1}).

\begin{figure}[!ht]
\centering
\includegraphics[width=0.5\textwidth]{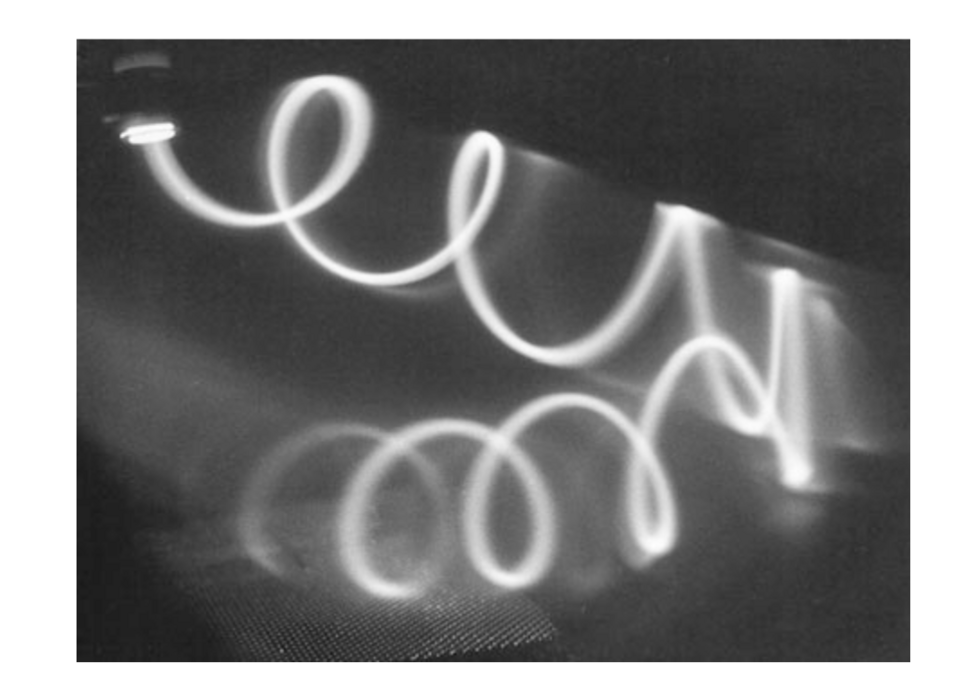} 
\setcaptionwidth{0.8\textwidth}
\caption{\footnotesize{Mirror reflection of an electron beam in a magnetic field that converges to the right. Note that the guiding center (axis of spiral) of the reflected beam does not coincide with that of the incident. This is due to the gradient and curvature drift in a nonuniform field. Courtesy of professor Reiner Stenzel. \cite{rein}}}
\label{mirror2}
\end{figure}

This phenomenon is sometimes called \textit{collisionless shock}, since the scattering occurs on a length scale which is smaller than the mean free path which characterize the particle collision. In fact, the scattering caused by magnetic mirrors (Fig. \ref{mirror2}) is due to the emission and the absorption of excitations of the plasma, called \textit{plasma waves}. In reality, these strong and inhomogeneous magnetic fields are due to frozen clouds of interstellar matter, whose density is much higher than the density of the surrounding material. \cite{chen}

\subsection{The Fermi acceleration model}

\begin{figure}[!ht]
\centering
\includegraphics[width=0.95\textwidth]{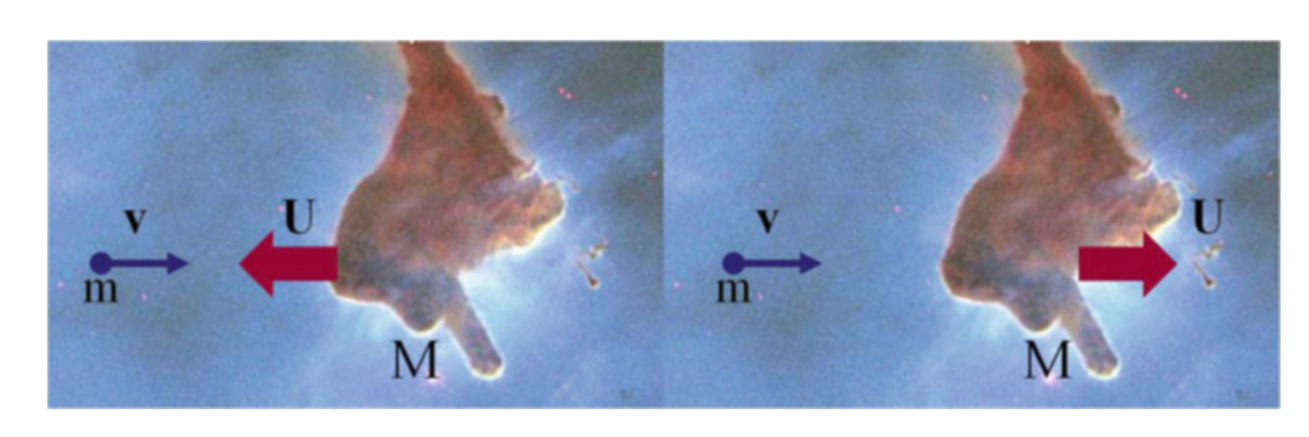} 
\setcaptionwidth{0.8\textwidth}
\caption{\footnotesize{On the left: the particle and the cloud velocities are opposite in direction. The particle gains energy in head-on elastic scattering. On the right: the particle and the cloud velocities are in the same direction. The particle loses energy in the elastic scattering.}}
\label{cloud}
\end{figure}

In 1949 E. Fermi suggested one of the first models for cosmic rays acceleration, which consists in particles accelerated by collisions with a moving cloud of gas. Let us assume that $\vec{v}$ and $\vec{U}$ are the initial velocities of a particle and a cloud respectively, with $\vec{v}\parallel \vec{U}$ and $v\gg U$; $m$ and $M$ are the masses of the particle and the cloud respectively, with $M\gg m$. Using energy-momentum conservation and assuming the scattering to be elastic, we easily obtain the magnitude of the velocity of the particle after the collision

\begin{equation}
    v'=-v\pm 2U\, ,
\end{equation}

\noindent where the sign depends on whether the velocities of the particle and the cloud are parallel or antiparallel (Fig. \ref{cloud}). Therefore the variation of the kinetic energy of the particle after each collision is

\begin{equation}
    \Delta E \simeq \pm \frac{4U}{v}E\, ,
\end{equation}

\noindent where we stopped at the first order in $\tfrac{U}{v}$. This means that in each collision the particle can gain or lose energy: considering a random distribution of particles moving in both directions and several collisions, on average a particle will gain energy, before escaping the acceleration region, emerging with a power-law distribution $N(E)=K\ E^{-\alpha}$.

This mechanism turns out to be very inefficient, especially for high-energy particles, so it was improved by Fermi himself, who in 1954 proposed that the acceleration is due to collisions between two clouds. Let us define two reference frames, $S$ and $S'$, which are the reference frame of the observer and the rest frame of a cloud respectively. We are also assuming that the velocity of the cloud has only $x$ component, so that only the $p_x$ component of the momentum of the particle is relevant (the others are conserved in the scattering). If the collision is elastic in $S'$, then the energy of the particle in this reference frame $E'$ is conserved after the collision, while momentum changes from $p_x$ to $-p_x$. Therefore we can calculate the variation of the particle energy in the observer reference frame, obtaining

\begin{equation}
    \Delta E=\frac{2\,v\,U\,\cos \theta}{c^2}+2\left(\frac{U}{c}\right)^2E\, ,
\end{equation}

\noindent where $\theta$ is the angle between $\vec{v}$ and the x-axis: energy is gained in head-on collisions ($\cos \theta >0$) and lost in catching collisions ($\cos \theta <0$). If we average the variation of the particle energy over all directions, the first term is null and the average energy variation is $\propto \left(\tfrac{U}{c}\right)^2$. This model is called \textit{second-order Fermi acceleration mechanism}. 

We can also consider an astrophysical situation in which only head-on collisions occur, e.g., two clouds mutually approaching or stellar material separated by a shock front: in these circumstances the second term in $\Delta E$ can be neglected (since $\tfrac{U}{c} \ll 1$) and for a relativistic particle $v\sim c$, so the energy gain becomes

\begin{equation}
    \Delta E=\frac{2U\cos \theta}{c} E\, ,
\end{equation}

\noindent while the particle energy after the scattering in the reference frame of the observer is

\begin{equation}
    E^*=E+\Delta E=\left(1+\frac{2U\cos \theta}{c}\right)E\, .
\end{equation}

\noindent The averages of $\Delta E$ and $E^*$ over all head-on directions ($\cos \theta >0$) are respectively

\begin{equation} \label{efficiency}
    \langle \Delta E \rangle=\frac{4U}{3c} \langle  E \rangle \equiv \eta \langle  E \rangle  \, ,
\end{equation}

\noindent and

\begin{equation}
    \langle  E^* \rangle=\left(1+\frac{4U}{3c}\right) \langle  E \rangle \equiv \mathcal{B} \langle  E \rangle  \, ,
\end{equation}

\noindent where the parameter $\eta \sim 10^{-2}$ represents the efficiency of the kinetic energy transfer from the accelerator to relativistic particles. This model, which considers only head-on collisions, is called \textit{first-order Fermi acceleration mechanism}. This is the mechanism behind particles acceleration in supernovae explosions. \cite{shosnr, sho3}

\subsection{The power-law energy spectrum from the Fermi model}

As we have seen at the end of the previous subsection, the Fermi model predicts that the particle gains energy after each collision, so that its energy becomes $ E = \mathcal{B}\, E_0 $ on average; therefore after $k$ collision its energy becomes $  E= \mathcal{B}^{\,k}\,  E_0 $. Since there is a probability $P$ that a particle remains in the acceleration region, after $k$ collision we expect that the number of particles with energy  $>E$ is

\begin{equation}
    N(>E)=P^{\,k}N_0 \, ,
\end{equation}

\noindent where $N_0$ is the initial number of particles in the acceleration region. Therefore it is trivial to show that the energy has a power-law spectrum:

\begin{equation}
    \frac{\ln \left(\frac{N(>E)}{N_0}\right)}{\ln \left(\frac{E}{E_0}\right)}=\frac{\ln P}{\ln \mathcal{B}}\equiv \alpha\, ,
\end{equation}

\noindent thus

\begin{equation}
    \frac{N(>E)}{N_0}=\left(\frac{E}{E_0}\right)^{\! \alpha} \, .
\end{equation}

\noindent The last equation can be written equivalently as

\begin{equation}
    \frac{dN}{dE}=N_0\left(\frac{E}{E_0}\right)^{\!\alpha-1} \, ,
\end{equation}

\noindent which is the power-law energy spectrum derived in the Fermi model. Moreover, it is possible to prove that $\alpha \approx -1$ via thermodynamic arguments, then we expect that the spectral index of the differential flux is $\tfrac{dN}{dE}\sim E^{-2}$ near sources. \cite{maxen}

\section{Supernova remnants and the Standard Model of cosmic rays acceleration}\label{quintasezione}

The scientific community agrees about the sources and the acceleration mechanism behind galactic cosmic rays: they originate in different types of supernova explosions and the particles are accelerated through diffusive transport in the neighbourhood of strong shock waves raised during the explosions. It is estimated that about 99\% of the energy released during a supernova explosion is emitted as neutrinos, while the remaining 1\% (which is $\sim 10^{51}$ erg if we consider a star whose mass is $10 \  \mathrm{M}_\odot$) is the kinetic energy of the expelled material, which forms the \textit{shock wave}, as we will see in the next subsection more in detail.

\subsection{Relevant quantities in supernova remnants}

Let us consider a supernova whose mass is $M=10 \  \mathrm{M}_\odot =2\cdot 10^{34} \ \mathrm{g}$ and its total binding (gravitational) energy is $E_g\simeq 2\cdot 10^{53}\ \mathrm{erg}$, which is therefore the total energy released per explosion. Since only the 1\% of this energy is transferred to the expelled material, the average energy emitted as kinetic energy is $K\simeq 2 \cdot 10^{51}$ erg. We can easily estimate the speed $U$ of the shock wave 

\begin{equation}
    U\simeq \sqrt{\frac{2K}{M}}=\sqrt{\frac{4\cdot 10^{51} \ \mathrm{erg}}{2\cdot 10^{34}\ \mathrm{g}}}\simeq 5 \cdot 10^8 \ \frac{\mathrm{cm}}{\mathrm{s}}\, ,
\end{equation}

\noindent and with respect to the speed of light

\begin{equation}
    \frac{U}{c}\simeq 2\cdot 10^{-2}\, ,
\end{equation}

\noindent which means that, even if it is a non-relativistic velocity, it is still larger than a typical velocity of interstellar materials. Moreover, some regions can reach even higher velocities, about $\tfrac{U}{c}\sim 10^{-1}$. The ratio between the speed $U$ and the speed of light is related to the \textit{efficiency} $\eta$ of the acceleration process (that we defined in Eq. \eqref{efficiency}). As the shock wave front expands through the interstellar material, it collects matter and decreases its velocity. When the mass of the swallowed matters becomes comparable to the mass of the ejected material, the velocity has decreased significantly and the shock wave is no longer efficient. Therefore we can estimate the radius within which the shock waves can accelerate particles

\begin{equation}
    R_{SN}=\left( \frac{3\cdot 10\  \mathrm{M}_\odot}{4\pi\, \rho_{ISM}} \right)^{\frac{1}{3}}=\left( \frac{6\cdot 10^{34}\  \mathrm{g}}{4\pi \cdot 1.6 \cdot 10^{-24} \ \mathrm{g\ cm}^{-3}} \right)^{\frac{1}{3}}=1.4\cdot 10^{19}\ \mathrm{cm}=5 \ \mathrm{pc}\, ,
\end{equation}

\noindent where we used the interstellar material density $\rho_{ISM}\simeq 6 \cdot 10^{-24} \ \mathrm{g\ cm}^{-3}$. Moreover, the duration of the shock wave is \cite{snr11, snr22}

\begin{equation}\label{duration}
    T_{SN}=\frac{R_{SN}}{U}=\frac{1.4\cdot 10^{19}\ \mathrm{cm}}{5 \cdot 10^8 \ \frac{\mathrm{cm}}{\mathrm{s}}}\simeq 3\cdot 10^{10} \ \mathrm{s} \simeq 1000 \ \mathrm{y}\, .
\end{equation}

\begin{figure}[!ht]
\centering
\includegraphics[width=0.6\textwidth]{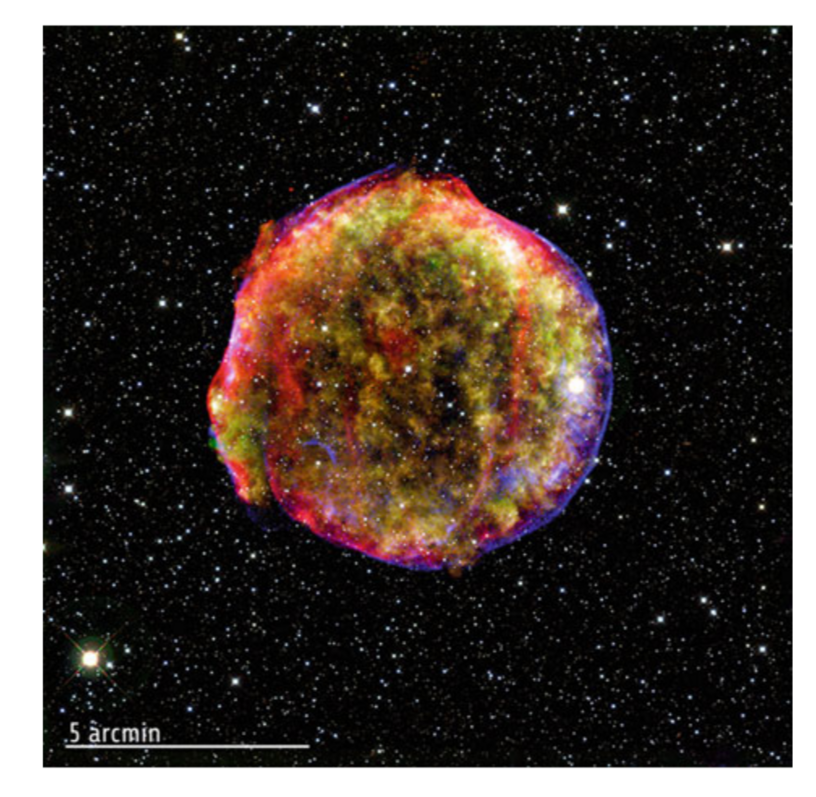} 
\setcaptionwidth{0.8\textwidth}
\caption{\footnotesize{This composite image of the Tycho Brahe supernova remnant combines X-ray and infrared observations obtained with NASA’s Chandra X-ray Observatory and Spitzer Space Telescope, respectively, and at the Calar Alto Observatory, Spain. \cite{chandra} It shows the scene more than four centuries after the brilliant star explosion witnessed by Tycho Brahe and other astronomers of that era. Credit: X-ray: NASA/ CXC/SAO, Infrared: NASA/JPL-Caltech; Optical: MPIA, Calar Alto, O.Krause et al.}}
\label{sn}
\end{figure}

\subsection{Maximum energy in the supernova model}

In order to determine the maximum energy attainable in the supernova model, it is useful to introduce the \textit{characteristic period} of the process $T_{cycle}$, defined as the time between two successive scattering processes. The accelerated particles are confined by magnetic fields, therefore their confinement region is the Larmor radius (Eq. \eqref{larmor})

\begin{equation}
    \lambda_{cycle}\equiv r_L=\frac{E}{Z\,e\,B} \, ,
\end{equation}

\noindent so the typical time between two successive scatterings in the rest frame of a shock wave front which is moving with a velocity $U$ is

\begin{equation}
    T_{cycle}= \frac{\lambda_{cycle}}{U}=\frac{E}{Z\,e\,B\,U} \, .
\end{equation}

\begin{figure}[!ht]
\centering
\includegraphics[width=0.6\textwidth]{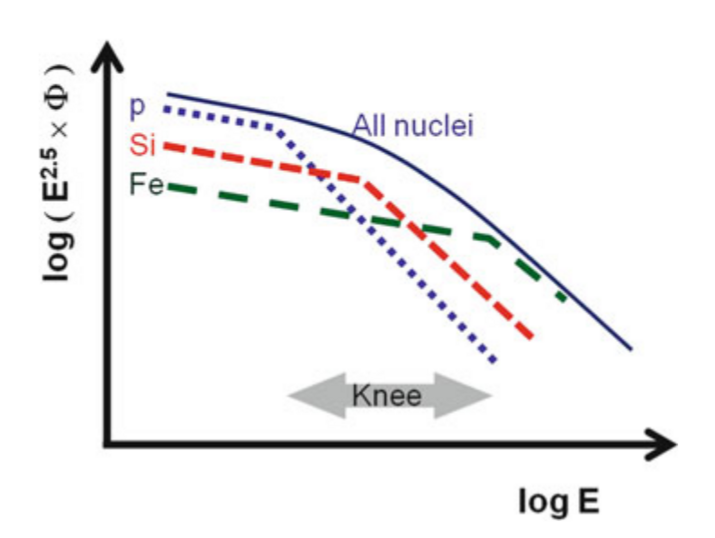} 
\setcaptionwidth{0.8\textwidth}
\caption{\footnotesize{The interpretation of the cosmic rays’ knee as being due to the correlation between the maximum energy and the atomic number $Z$. The flux of each nuclear species sharply decreases after a given cut-off, which depends on $Z$ as $E_{max}^Z\equiv E_{max}^p \cdot Z$, where $E_{max}^p$ is the maximum energy reached by protons. The behavior of hydrogen ($Z=1$), silicon ($Z = 14$) and iron ($Z = 26$) nuclei are depicted in figure.}}
\label{kneem}
\end{figure}

\noindent In each scattering, the energy of the particle increases at a rate which is given by the ratio between Eq. \eqref{efficiency} and the characteristic period

\begin{equation}
    \frac{dE}{dt}\simeq\frac{\eta \,E}{T_{cycle}}=\frac{\eta\, E\,Z\,e\,B\,U}{E}=\eta\, Z\,e\,B\,U\, .
\end{equation}

\noindent As we can see, the rate of energy gain does not depend on the particle energy $E$. Thus, the maximum energy attainable in this model is

\begin{equation}
    E_{max}\simeq T_{SN}\,\frac{dE}{dt}=\frac{R_{SN}}{U}\eta\, Z\,e\,B\,U\simeq\frac{U}{c}Z\,e\,B\, ,
\end{equation}

\noindent where we used $\eta\simeq\tfrac{U}{c}$ and the equation \eqref{duration} for $T_{SN}$. If we substitute the numerical values of the quantities involved, i.e., $\tfrac{U}{c}\simeq 2\cdot 10^{-2}$, $e=4.8\cdot 10^{-10}\ \mathrm{esu}$, $B\simeq 4\cdot 10^{-6} \ \mathrm{G}$, we obtain $E_{max}\simeq 500 \cdot Z \ \mathrm{erg} \simeq 300\cdot Z\ \mathrm{TeV}$, which depends on the atomic number of the particle. This means that this mechanism allows to reach energies of hundreds of TeV, which correspond to the energy region around the knee (Fig. \ref{kneem}). \cite{shosnr2}

\subsection{Energy considerations on cosmic rays sources}

As we have discussed in the previous sections, a magnetic field confines charged particles and so do galactic magnetic fields with cosmic rays: these can confine particles for a certain period of time $\tau_{esc}$ before they escape, which is estimated to be about $10^7$ y when averaged over all cosmic ray energies. This value decreases as cosmic ray energies increase; it would be about $10^5$ y at 100 TeV. The \textit{escape} or \textit{confinement time} $\tau_{esc}$ is defined as the average time required for a particle to reach the galactic boundary, outside of which the magnetic fields are negligible. Being $\rho_{CR}(>E_0)$ the energy density of cosmic rays whose energy is above $E_0$, the total kinetic energy of cosmic rays confined inside the galactic volume $\mathcal{V}_G\simeq 5\cdot 10^{66} \ \mathrm{cm}^3$ is then $\rho_{CR}(>E_0)\,\mathcal{V}_G$. This is about $8\cdot 10^{54}$ erg for $E_0=100$ TeV. We would expect that this energy increases over time in the presence of new galactic sources as the cosmic rays are confined inside the Galaxy, but we have to consider also that some cosmic rays escape from the Galaxy with the characteristic time $\tau_{esc}$. Therefore, the rate of energy loss is

\begin{equation}
    P_{CR}=\frac{\rho_{CR}(>E_0)\,\mathcal{V}_G}{\tau_{esc}}\simeq 3\cdot 10^{40} \ \frac{\mathrm{erg}}{\mathrm{s}}\, .
\end{equation}

\noindent This is then the power required by cosmic accelerators in order to keep in a stationary condition the energy density of cosmic rays in the galactic volume. Notice that here we are (reasonably) assuming that $\rho_{CR}\sim \mathrm{constant}$ for a time scale much bigger than the confinement time.

If we consider that the supernova rate in a galaxy like the Milky Way is about $2\div3$ per century (the characteristic time of this event is then $\tau_{SN}\simeq 10^9$ s) and that each supernova explosion emits on average $K\simeq 2\cdot 10^{51}$ erg as kinetic energy, then a physical process that accelerates particles transferring energy from the kinetic energy of the shock waves with an efficiency $\eta$ has the power

\begin{equation}
    P_{SN}=\eta\frac{K}{\tau_{SN}}\simeq \eta \cdot 2\cdot 10^{42} \ \frac{\mathrm{erg}}{s} \, .
\end{equation}

\noindent Requiring that $P_{SN}=P_{CR}$. we obtain that the efficiency of this process should be $\eta\sim 10^{-2}$, witch is the same efficiency provided by the Fermi model. \cite{mamasnr}

\subsection{Energy spectrum of cosmic rays at sources}

The loss probability of the cosmic ray confinement is energy dependent 

\begin{equation}
    P(E)=\left(\frac{E}{E_0}\right)^{\!-\delta}\, ,
\end{equation}

\noindent where $\delta$ is a parameter to be determined experimentally, fitting data on the base of a particular theoretical model: in most models $\delta\simeq 0.4\div 0.6$.

As we have seen in Sec. \ref{spectra}, below the knee the energy spectrum can be approximated by a power low, $\Phi(E)\sim E^{-\alpha}$, where $\alpha\simeq 2.7$ is the spectral index. Therefore we can calculate the energy spectrum at sources $Q(E)$ as

\begin{equation}
    \Phi(E)=Q(E)\,P(E)\sim Q(E)\, E^{-\delta} \, ,
\end{equation}

\noindent thus

\begin{equation}
    Q(E)\sim \Phi(E)\, E^{\delta}\sim E^{\delta-\alpha}=E^{-(2.1\div 2.3)} \, .
\end{equation}

\noindent This provide a very important information, because it says that a cosmic rays source should reproduce an energy dependence like this one, i.e., a power law with a spectral index close to 2. \cite{webber}


\subsection{Neutron stars and pulsars}

In the intermediate region of energy range between $10^{16\div 19}$ eV, the different possible sources, galactic or extragalactic, are compact objects able to accelerate particles \cite{2004hpa..book.....L,2008Sci...322.1221A}:




\begin{itemize}
    \item Neutron stars: a neutron star is an object made up by neutrons, with a radius of about $R\simeq 10$ km, produced by the collapse of massive stars ($8\div 25$ $\mathrm{M}_\odot$) after the supernova explosion, its mass is about   $1.4 \ \mathrm{M}_\odot$ and it has a high magnetic field of the order of $B_{NS}\simeq 10^{11\div 12}$ G on its surface. By the conservation of the angular momentum, neutron stars can rotate at very high speed ($\omega\simeq 10^4$ rad $\mathrm{s}^{-1}$);
    
    \item Pulsars: a pulsar is a rotating neutron star emitting a beam of electromagnetic radiation along its magnetic axes.
\end{itemize}

In order to accelerate a particle we need a variable magnetic field $B$ able to produce an induced electric field $\mathcal{E}$ on a length scale $L$ through the Maxwell-Faraday's law (in CGS units):

\begin{equation}
    \frac{\mathcal{E}}{L}=\frac{dB}{c\,dt}\label{mfl} \, ,
\end{equation}

\noindent where $c$ is the speed of light. The maximum energy gained by this mechanism in Eq. \eqref{mfl} is given by

\begin{equation}\label{hillasc}
    E_{max}= \int\! dx \ \  Z\,e\,\mathcal{E}=\int\! dx \ \  Z\,e\,L\frac{dB}{c\,dt}=\int\! dB \ \ Z\,e\,L\frac{dx}{c\,dt}=Z\,e\,L\,B\,\beta \, ,
\end{equation}

\noindent that is called the \textit{Hillas condition}, where $\tfrac{dx}{dt}$ is the velocity of the particle. For instance, if we have a pre-accelerated particle by this shock wave mechanism produced by the supernova explosion, passing close to the neutron star ($L\simeq R_{NS}$) with velocity $\beta\, c=\omega_{\,NS}R_{NS}=0.1\ c$, with $Z=1$ and $B\simeq10^{11}$ G, the maximum energy will be of the order of

\begin{equation}
    E_{max}=Z\,e\,R_{NS}\,B\,\frac{\omega_{\,NS}R_{NS}}{c}=5\cdot 10^{6}\  \mathrm{erg}\simeq 10^{18}\  \mathrm{eV}\, .
\end{equation}

\begin{figure}[!ht]
\centering
\includegraphics[width=0.6\textwidth]{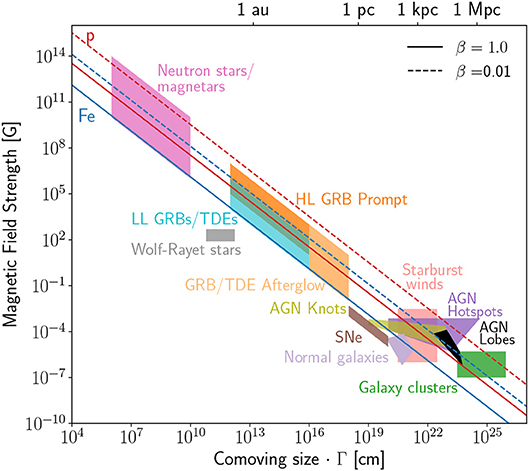} 
\setcaptionwidth{0.8\textwidth}
\caption{\footnotesize{Hillas diagram \cite{2019FrASS...6...23B}.
}}
\label{hillasd}
\end{figure}

\noindent The Hillas condition puts the maximum energy gained by a particle in relation to the length scale of the system in which we have acceleration and the magnetic field. There are many theoretical mechanisms which use compact objects in order to explain the maximum energy of cosmic rays up to $10^{18}$ eV but there are no models using galactic objects which can reach up to $10^{20}$ eV.


The Hillas diagram in Fig. \ref{hillasd} shows classes of objects in terms of the product of their radial size $R$, magnetic field $B$, and associated uncertainty in the ideal Bohm limit where $\eta=1$ ($\eta$ is the efficiency of acceleration).

\subsection{Possible galactic cosmic ray sources above the knee}

The condition of Eq. (6.10) defines the fact that the power emitted by Galactic sources in terms of relativistic protons and nuclei (i.e., kinetic energy larger than $\sim$ few GeV) should be of about $ 10^{40}$ erg/s.  Supernovae explosions in our Galaxy are natural candidates for providing such a power supply. 



The maximum energy that seems reachable by protons and nuclei accelerated by the supernova shock wave mechanism could be in the range of $10^6-10^7$ GeV.  To describe the steady presence of cosmic rays of energy larger than 1 PeV= $10^6$ GeV (the region in the cosmic ray spectrum above the knee), we need sources accelerating cosmic rays with a power $P(>1)$ PeV = $10^{37-38}$ $\mathrm{s}^{-1}$.



\begin{figure}[!ht]
\centering
\includegraphics[width=0.8\textwidth]{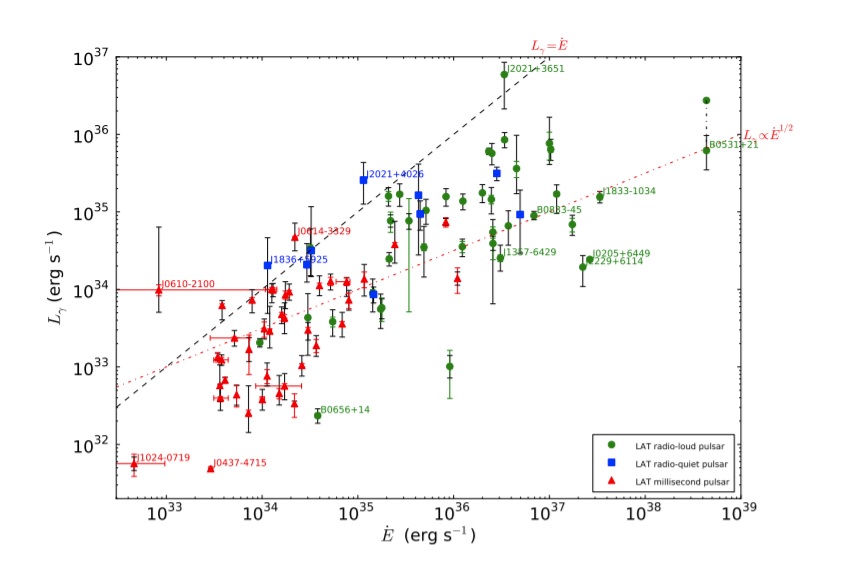} 
\setcaptionwidth{0.8\textwidth}
\caption{\footnotesize{$\gamma$-ray luminosity $L_\gamma$ (erg $\mathrm{s}^{-1}$) in the 0.1 to 100 GeV energy band versus the total energy loss $\Dot{E}$ (erg $\mathrm{s}^{-1}$) \cite{2013ApJS..208...17A}.
}}
\label{lum}
\end{figure}

The largest fraction of galactic objects emitting $\gamma$-rays observed by the Fermi-LAT collaboration are pulsars and in Fig. \ref{lum} it is represented the luminosity in $\gamma$-ray emission $L_\gamma$ vs the total energy loss $\Dot{E}$. The variation of the periodicity of the pulsar is a measure of the total energy loss. Pulsars work for a limited period of time: with the energy loss there is a decrease of the frequency and with a slow rotation any acceleration mechanism is canceled.

In conclusion, cosmic rays produced with energy below $10^{15}$ eV are the convolution of many supernova explosions during the history of our Galaxy in the last 10 million years, but to explain cosmic rays in the region between the knee and the ankle ($10^{15\div 18}$ eV) we need an estimate of order of 10$\div$100 sources which emit accelerated charged particles which reach the Earth, some of them producing $\gamma$-rays. On the other side, the observation of $\gamma$-rays it is not an indication of a detected source of cosmic rays because they are produced either by electrons or protons  \cite{2012APh....39...52D,bra,los,1990cup..book.....G,1990acr..book.....B}.






\section{Acceleration of cosmic rays from extragalactic sources} \label{sestasezione}

Cosmic rays with energies above $10^{19}$ eV are usually referred as Ultra-High Energy Cosmic Rays (UHECRs). The motivation for an extragalactic origin of ultra-high energy cosmic rays is shown in the right panel in Fig. \ref{galactic}: galactic magnetic fields have small effects on protons above $10^{19}$ eV which move in straight way, and a galactic source would manifest itself through an anisotropy from the galactic plane which is unobserved.




The travel distances of particles coming off the Milky Way are very high and during the propagation we have three main energy loss processes for protons and heavy nuclei:

\begin{itemize}
    \item the adiabatic energy loss \cite{bra} due to the expansion of the Universe: we first define the energy loss length $l$ as 
    \begin{equation}
        l^{-1}\equiv\frac{1}{E}\frac{dE}{dx} \, ,
    \end{equation}
    
    and the adiabatic loss of a cosmic ray with energy $E$ is
    \begin{equation}
        -\frac{1}{E}\frac{dE}{dx}=H_0\, ,
    \end{equation}
    thus
     \begin{equation}
        l_{adia}=\frac{c}{H_0}\simeq4 \ \mathrm{Gpc} \, ;
    \end{equation}
  
\item pion-production on photons of the CMB (Cosmic Microwave Background) radiation: ultra-high energy cosmic rays can suffer interaction with the CMB radiation due to the process of production of $\Delta^+$ which decays in pions and neutrons \cite{1969PhLB...28..423B} :

\begin{equation}
    \begin{aligned}\label{pions}
        p+\gamma_{CMB}\to\Delta^+\to & \pi^++n \\
        \to & \pi^0+p \, .
    \end{aligned}
\end{equation}

\noindent The $\pi^0$ decays in two photons, while the $\pi^+$ decays into $\mu^+\nu_{\mu}$. The proton loses 10\% of the initial energy in each process; the neutron is not observed because it decays after 15 minutes in a proton that in the final step has a reduced energy.
The process in Eq.\eqref{pions} occurs if the proton has energy larger than $E_{GZK}\simeq5\cdot10^{19}$ eV that is the \textit{GZK cut-off} and, for a nucleus of mass $A$, for energy larger than $A E_{GZK}$. 

The energy loss length $l_{p\gamma}$ of a proton in the CMB radiation for this process is

\begin{equation}
l_{p\gamma}=\frac{1}{\bigl\langle y\sigma_{p\gamma}n_{p\gamma}\bigr\rangle}=30 \ \mathrm{Mpc}\, ,
    \end{equation}
where $y=\frac{E-E'}{E}$ is the fraction of energy lost per interaction, $n_{p\gamma}$ is the corresponding photon number density and the brackets remind us that we should integrate
the differential cross-section $\sigma_{p\gamma}$ over the momentum distribution of the target photons \cite{2000CoPhC.124..290M}. If we observe a very large number of protons at energies larger that the GZK cut-off they probably originated in a region smaller than 30 Mpc;

\item electron-positron pair-production on the CMB radiation: during the propagation in the CMB radiation, $e^+e^-$ pairs can be also produced in the process \cite{2012APh....39..129A}:
\begin{equation}
    p+\gamma\to p+e^+e^- \, .
\end{equation}
\end{itemize}

\begin{figure}[!ht]
\centering
\includegraphics[width=0.7\textwidth]{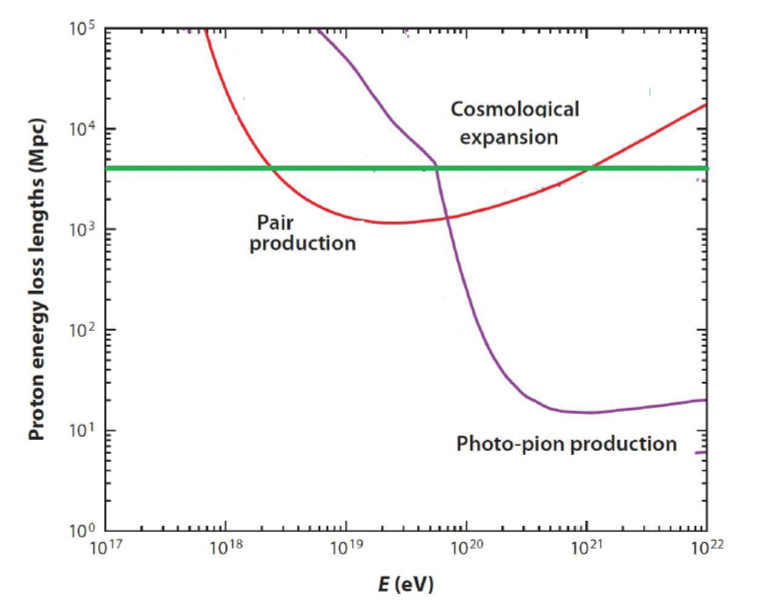}
\setcaptionwidth{0.8\textwidth}
\caption{\footnotesize{The energy loss lengths for a high-energy proton propagating through the CMB radiation field. Pair creation, photo-pion production, and energy loss through cosmological expansion are shown.}}
\label{scale}
\end{figure}

\noindent In Fig. \ref{scale} we can see the variation of the energy loss scale length as a function of the energy for the three considered processes.



\begin{figure}[!ht]
\centering
\includegraphics[width=0.55\textwidth]{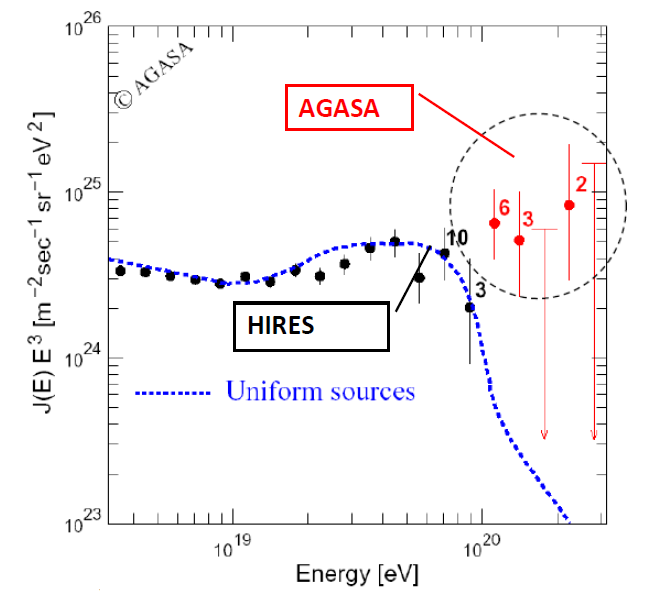}
\setcaptionwidth{0.8\textwidth}
\caption{\footnotesize{AGASA and HiRes' results in the 90s.}}
\label{aga}
\end{figure}

During the 90s two experiments were led: AGASA (Akeno Giant Air Shower Array) in Japan and HiRes (High Resolution Fly's Eye) in the USA. The results of the two experiments were incompatible because AGASA observed 11 events above $10^{20}$ eV while HiRes detected no events of this type (see Fig. \ref{aga}) \cite{2008APh....30..167K,2008PhRvL.100j1101A,2003APh....19..447T}. AGASA's results suggested the idea of the \textit{top-down model}, i.e., the existence of massive particles whose decay could produce the observed cosmic rays able to violate the GZK model. This model does not accelerate charged particles (“\textit{bottom-up model”}) and it uses the stable secondaries produced in the fragmentation of superheavy particles. Two main variants of top-down models, well motivated independently of ultra-high energy cosmic rays observations, were developed:

\begin{itemize}
    \item superheavy dark matter;
    \item topological defects.
\end{itemize}

\noindent Direct implications of top-down models are both the non-existence of a cosmic rays suppression and the prediction of a large flux of ultra-high energy $\gamma$-rays and neutrinos.

\begin{figure}[!ht]
\centering
\includegraphics[width=0.65\textwidth]{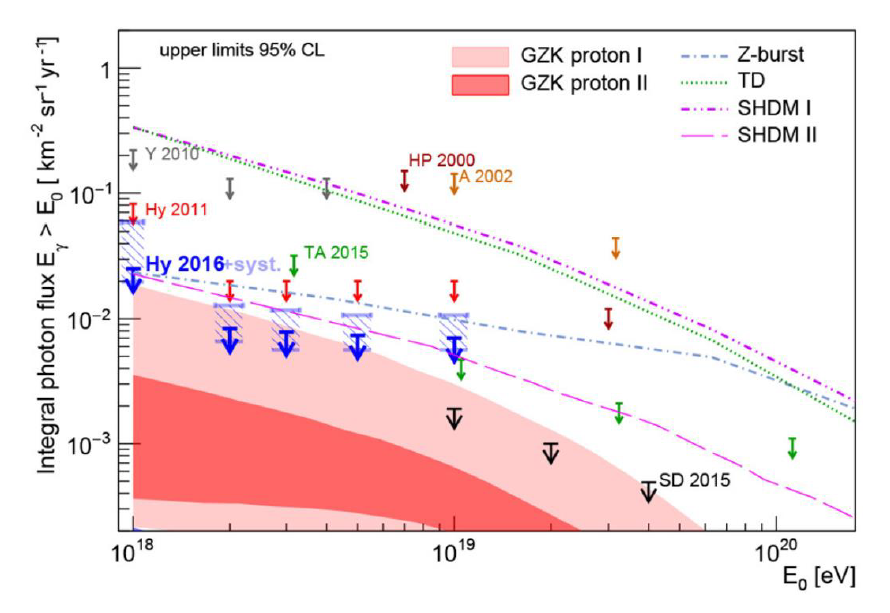}
\setcaptionwidth{0.8\textwidth}
\caption{\footnotesize{Blue arrows: upper limits from the AUGER 9 year hybrid data sample; black arrows: 9 year SD data; other symbols: previous data from AUGER as well as data from TA, AGASA, Yakutsk, and Haverah Park; lines: predictions for top-down models; shaded regions: expected from GZK photon fluxes, assuming different parameters. Auger \cite{Aab_2017}.}}
\label{ph}
\end{figure}

\noindent Starting from 2005 new experiments were led using hybrids detector (e.g., the PAO, Pierre Auger Observatory) \cite{2017ApJ...850L..35A,2017arXiv170806592T,2017Sci...357.1266P,2015PhRvD..91i2008A,2008PhRvL.101f1101A} which measured the flux of ultra-high energy photons, but the large flux of $\gamma$-rays and neutrinos was not observed. The experimental results are illustrated in Fig. \ref{ph}. These observational facts have strongly reduced interest in top-down models.

The \textit{bottom-up models} are candidates as astrophysical sources, we can distinguish among \cite{2002ApJ...579..530W,2011ARA&A..49..119K}:

\begin{itemize}
    \item Galactic sources such as
    \begin{itemize}
        \item \textit{Micro-quasars}, a compact object (black hole or neutron star) towards which a companion star is accreting matter. Neutrino beams could be produced in the micro-quasar jets;
        \item \textit{Supernova remnants}, several different objects (with different neutrino production scenarios) such as plerions (center-filled supernova remnants), shell-type supernova remnants and supernova remnants with energetic pulsars;
        \item \textit{Magnetars}, isolated neutron stars with surface dipole magnetic fields $\sim10^{15}$ G, much larger than ordinary pulsars. A seismic activity in the surface could induce particle acceleration in the magnetosphere.
    \end{itemize}
    \begin{figure}[!ht]
    \centering
    \includegraphics[height=0.26\textwidth]{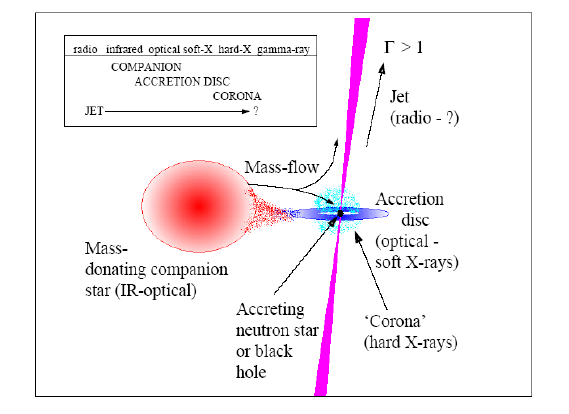}
    \includegraphics[height=0.26\textwidth]{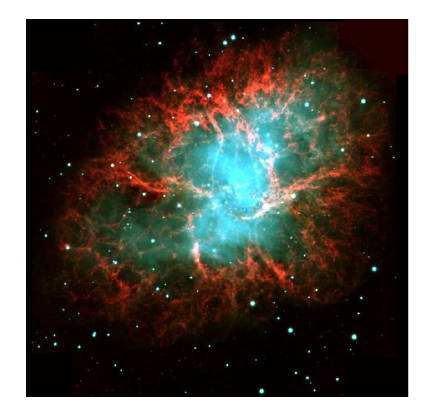}
    \includegraphics[height=0.26\textwidth]{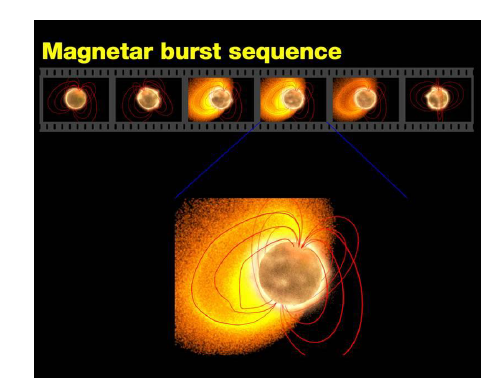}
    \setcaptionwidth{0.8\textwidth}
    \caption{\footnotesize{Micro-quasar scheme (left panel), supernova remnant (Crab Nebula, center), magnetar (right panel).}}
    \label{micro}
    \end{figure}
    \begin{figure}[!ht]
    \centering
    \includegraphics[height=0.30\textwidth]{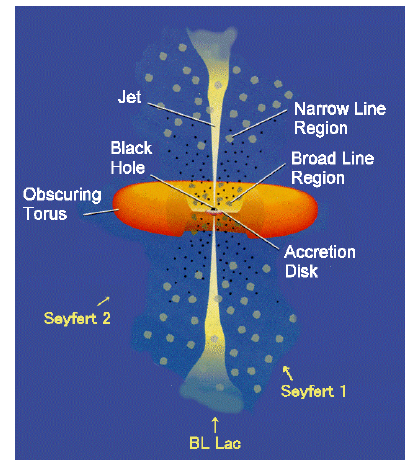}
    \includegraphics[height=0.30\textwidth]{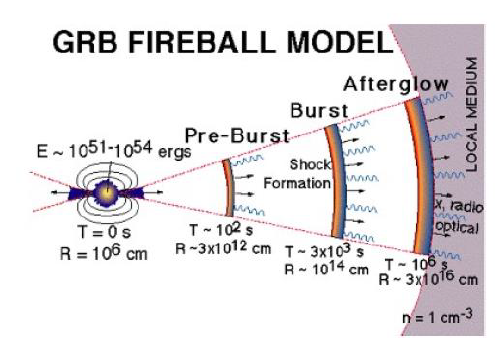}
    \setcaptionwidth{0.8\textwidth}
    \caption{\footnotesize{Active galactic nucleus scheme (left panel), $\gamma$-ray burst fireball model (right panel).}}
    \label{agn}
    \end{figure}
    \item Extra-galactic sources such as
    \begin{itemize}
        \item \textit{Active galactic nuclei}, includes Seyferts, quasars, radio galaxies and blazars. For the standard model: a super-massive ($10^{6\div 9}$ $\mathrm{M}_\odot$) black hole towards which large amounts of matter are accreted;
        \item \textit{Gamma-ray bursts}, brief explosions of $\gamma$-rays. In the fireball model, matter moving at relativistic velocities collides with the surrounding material. The progenitor could be a collapsing super-massive star. The time correlation enhances the neutrino detection efficiency.
    \end{itemize}
\end{itemize}

\section{Secondary particles: $\gamma$-rays and neutrinos} \label{settimasezione}

\begin{figure}[!ht]
    \centering
    \includegraphics[width=0.7\textwidth]{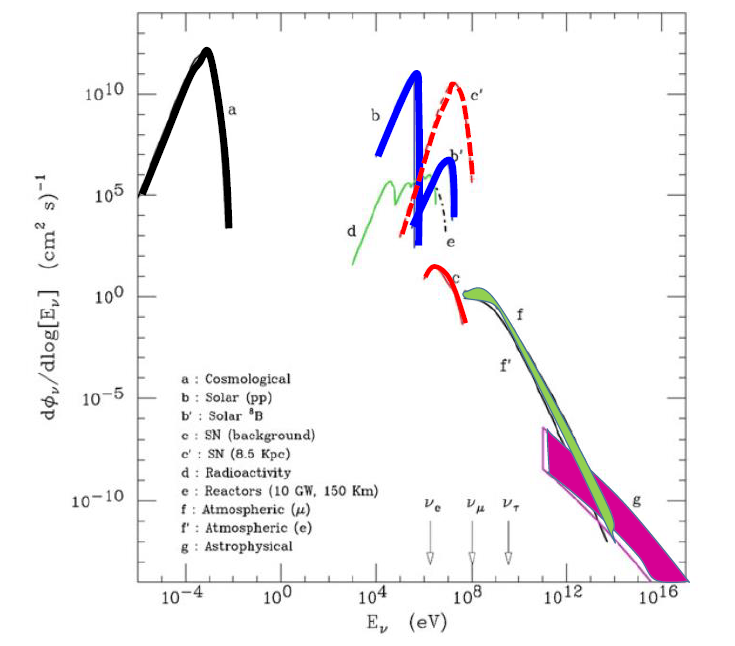}
    \setcaptionwidth{0.8\textwidth}
    \caption{\footnotesize{Flux of neutrinos at the surface of the Earth. Arrows indicate the energy threshold for charged current production of the charged lepton. Black line: Big Bang neutrinos; blue line: neutrinos from the Sun; red line: neutrinos from supernovae; green line: atmospheric neutrinos; magenta line: high-energy cosmic neutrinos.}}
    \label{neu}
\end{figure}

In the near future one of the probes which can be important to distinguish between models, to improve models and to point to the direction of some cosmic rays accelerators, is the neutrino. Neutrinos are neutral particles and their limited cross sections reduce their absorption by matter; the construction of detectors able to catch neutrinos is a hard job. Neutrinos' spectrum is given by the plot in Fig. \ref{neu} in which we have the flux of particles for square centimeter per second as a function of the energy $E$, for neutrinos of different kinds.

\begin{figure}[!ht]
    \centering
    \includegraphics[width=0.6\textwidth]{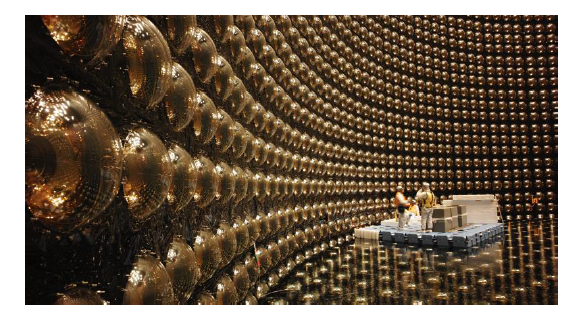}
    \setcaptionwidth{0.8\textwidth}
    \caption{\footnotesize{Super-Kamiokande experiment.}}
    \label{ka}
\end{figure}

There are different kind of neutrinos: an expected but still not measured flux of cosmological neutrinos (Big Bang neutrinos), solar neutrinos (a milestone for multi-messengers astrophysics), neutrinos from supernovae (an observation in 1987, the only supernova occurring closer to the solar system from the epoch in which solar neutrinos detector were ready), atmospheric neutrinos (whose discovery by Super-Kamiokande \cite{2012NuPhS.229...79T,2007PhRvD..75d3006H}, see Fig. \ref{ka}, induced a modification of the standard model) and high energy cosmic neutrinos produced by galactic and extra-galactic acceleration.


\begin{figure}[!ht]
    \centering
    \includegraphics[width=0.82\textwidth]{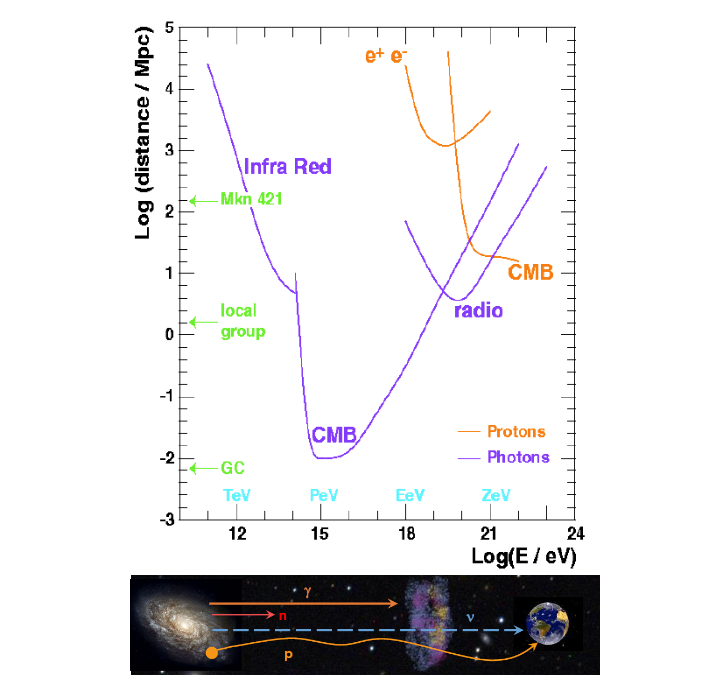}
    \setcaptionwidth{0.8\textwidth}
    \caption{\footnotesize{The photon and proton mean free range path and the interaction with matter and radiation.}}
    \label{pp}
\end{figure}

Neutrinos are important because they do not suffer absorption as $\gamma$-rays while photons can interact with CMB radiation ($r\simeq10$ kpc, $E\simeq100$ TeV: above hundred TeV it is unlikely that they can arrive from extra galactic source) and other radiation fields and matter; neutrinos are not subject to deflection by magnetic fields as the protons which interact with CMB ($r\simeq10$ Mpc, $E\simeq10^{11}$ GeV), while neutrons are not stable. In Fig. \ref{pp} it is shown the photon and proton mean free range path and the interaction with matter and radiation.

\begin{figure}[!ht]
    \centering
    \includegraphics[width=0.7\textwidth]{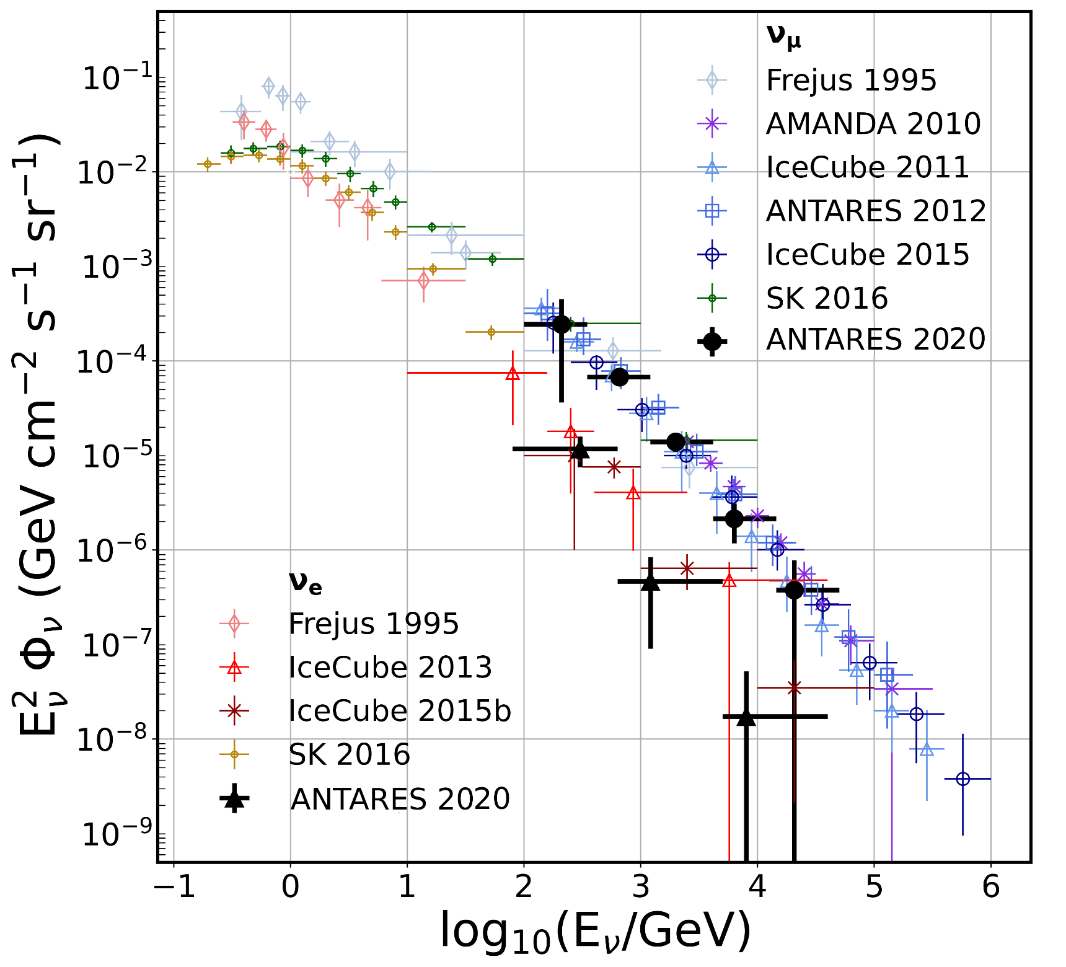}
    \setcaptionwidth{0.8\textwidth}
    \caption{\footnotesize{Measured energy spectra of the atmospheric $\nu_e$ and $\nu_\mu$ using shower-like and starting track events in the ANTARES neutrino telescope (black). The measurements by other experiments (Frejus, AMANDAII, IceCube, and Super-Kamiokande), as well as the previous $\nu_\mu$ flux measurement using a different ANTARES data sample, are also reported. The vertical error bars include all statistical and systematic uncertainties.}}
    \label{nn}
\end{figure}

The idea of a large volume experiment for cosmic neutrinos based on the detection of the secondary particles produced in neutrino interactions was first formulated in the 1960s by M. Markov that proposed to install detectors deep in a lake or in the sea and to determine the direction of the charged particles with the help of Cherenkov radiation. At present a $\mathrm{km}^3$ detector (IceCube \cite{2013Sci...342E...1I,2013ApJ...779..132A,2016ApJ...833....3A,2017arXiv171001191I,2006EPJC...46..669H}) is operating in the ice of the South Pole and another smaller underwater telescope (ANTARES \cite{2012ApJ...760...53A,2013A&A...559A...9A,2014ApJ...786L...5A,2018ApJ...853L...7A}) is running in the Mediterranean Sea, waiting for the Mediterranean $\mathrm{km}^3$ telescope (KM3NeT \cite{2016JPhG...43h4001A}) and another detector in Lake Baikal. All of them use the Markov idea and are made up of a grid of optical sensors (photomultipliers, PMTs \cite{2008NIMPA.597...23A}) inside the so-called instrumented volume. In Fig. \ref{nn} is shown the measured energy spectra of the atmospheric $\nu_e$ and $\nu_\mu$ in the different experiments.

The astrophysical production of high-energy neutrinos occurs via the decay of charged pions in the beam dump of energetic protons in dense matter

\begin{equation}
    p + p\to\pi^\pm,\pi^0,K^\pm,K^0,p,n,\dots \, ,
    \label{bd}
\end{equation}

\noindent where "$\dots$" represents the presence of higher mass mesons and baryons. The $\pi^0$ decays immediately in two $\gamma$-rays

\begin{equation}
    \pi^0\to\gamma+\gamma\, ,
    \label{zero}
\end{equation}

\noindent the charged pions decay as 

\begin{equation}
    \begin{split}
        \pi^-\to\mu^-+\Bar{\nu_\mu} \, ,\\
        \pi^+\to\mu^++\nu_\mu \, ,
    \end{split}
    \label{pio}
\end{equation}

\noindent and in turn, there is the decay
    
\begin{equation}
    \begin{split}
        \mu^-\to e^-+\Bar{\nu_e}+\nu_\mu \, ,\\
        \mu^+\to e^++\nu_e+\Bar{\nu_\mu}\, .
    \end{split}
    \label{muo}
\end{equation} 

\noindent Thus, there are three neutrinos for each pion, and six neutrinos for every two $\gamma$-rays. The production of high-energy neutrinos also occurs through photoproduction hadronic processes from cosmic ray protons interacting with ambient photons as seen in Eq. \eqref{pions}.

\subsection{Neutrino detection}

A neutrino telescope measures the neutrino interactions in the unit of time, so we have the event rate ($\mathrm{s}^{-1}$) at time $T$. Neutrinos produced by astrophysical sources go toward the detector 
and interact with water producing a muon above 1 TeV according to Eq. \eqref{muo} and 
the muon follows the same path of neutrino: 
the light produced by the passage of the muon is measured or, if the neutrinos interact close to the detector, there is the detection of the cascade produced by the particles.

\begin{figure}[!ht]
    \centering
    \includegraphics[width=0.7\textwidth]{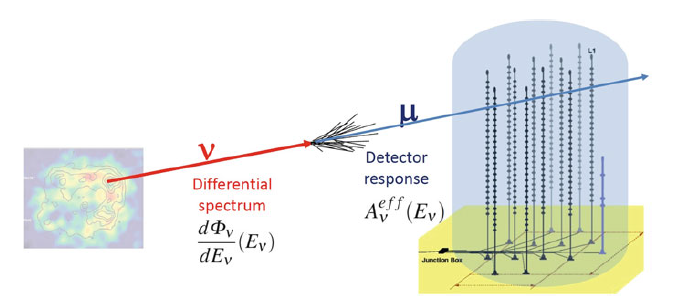}
    \setcaptionwidth{0.8\textwidth}
    \caption{\footnotesize{The number of observed events in a large neutrino detector is given by the convolution of the differential neutrino flux and the effective neutrino area of the detector.}}
    \label{det}
\end{figure}

We have astrophysical information about the neutrino flux and the response of the detector is included in the \textit{neutrino effective area} $A_\nu^{eff}(E)$, an experimental parameter which depends on the energy and on the incoming direction of the particle, on the status of the detector, on the cuts that each particular analysis uses for the suppression of the background, on the neutrino flavor and on the particular kind of signal we are looking for (events induced by muons or by a cascade produced by neutral current/charged current interactions). Only detailed and dedicated Monte Carlo simulations can determine the neutrino effective area. The relation between the neutrino effective area $A_\nu^{eff}(E)$ and the event rate $N_\nu$ at time $T$ is \cite{2010EPJC...65..649C,2019arXiv191009065N}

\begin{equation}
    \frac{N_\nu}{T}= dE\,\frac{d\phi_\nu}{dE}\,A_\nu^{eff}(E)\, ,
    \label{eventrate}
\end{equation}

\noindent where $\frac{d\phi_\nu}{dE}$ is the neutrino flux. The effective area
corresponds to the quantity that, convoluted with the neutrino flux, gives the event
rate as seen in Fig. \ref{det}.

\begin{figure}[!ht]
    \centering
    \includegraphics[width=0.7\textwidth]{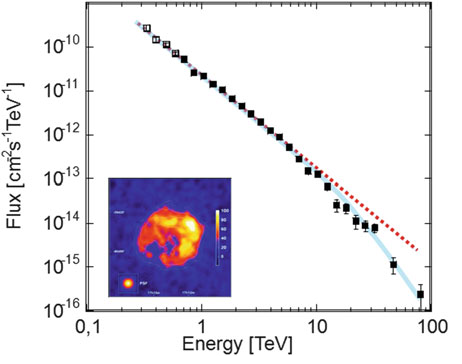}
    \setcaptionwidth{0.8\textwidth}
    \caption{\footnotesize{HESS Measurement of the $\gamma$-ray flux from RX J1713.7-3946 (SNr).}}
    \label{flux}
\end{figure}

If the hadronic mechanism is at work in a galactic source (for instance, in a supernova remnant accelerating cosmic rays), a flux of neutrinos comparable to the one of $\gamma$-rays could be expected. It is useful to define a sort of reference neutrino flux from a galactic source.
The supernova remnant RX J1713.7-3946 has been the subject of large debates about the nature of the process (leptonic or hadronic) that originates its $\gamma$-ray spectrum.
This source has been observed with high statistics by the HESS telescope up to $∼80$ TeV, with a spectrum that can be reasonably well described by a power law with an exponential cut-off:

\begin{equation}
    E^2\,\frac{d\phi_\nu}{dE}=\phi_0^\gamma\, e^{-\sqrt{\textstyle{\frac{E}{E_c}}}} \, ,
\end{equation}

\noindent where $\phi_0^\gamma= 1.8 \cdot 10^{−11}$ TeV $\textrm{s}^{−1}$ $\textrm{cm}^{−2}$ and the cut-off parameter $E_c=3.7$ TeV, see  Fig. \ref{flux}.

Different models exist giving predictions about the neutrino flux; the expected number of $\nu_\mu$ is equal to that of $\gamma$-rays, as in proton-proton interactions in which the same number pions are produced. Two photons arise from the $\pi^0$ decay, and six neutrinos from the decay chains of the two charged pions. At a large distance from the source, two out of six arrive as $\nu_\mu$ when neutrino oscillations are considered. However, the flux of high-energy neutrinos is about a factor lower with respect to the $\gamma$-rays of the same energy, because of the kinematics. In the decay of charged pions, a larger fraction of kinetic energy is transferred to the muon: in the pion rest frame $∼110$ MeV to the muon and $∼30$ MeV to the neutrino. The results of the predictions agree on the fact that the neutrino flux is, at first order

\begin{equation}
    \frac{d\phi_\nu}{dE}\simeq\frac{d\phi_\gamma}{dE}=10^{-11}\ E^{-2}\ \mathrm{TeV}\ \mathrm{cm}^{-2}\ \mathrm{s}^{-1} \, .
\end{equation}

In a neutrino detector we can observe two main kind of interactions: events with a long track due to a passing muon, and events with a shower, without the presence of a muon.  In the track-like events the $\nu_\mu$ produced by charged current interactions produce a muon:
the neutrino interactions occurs 10 km from the detector while the muon at 1 TeV can reach the detector;  what is measured is the muon energy  and in this case there is an underestimation of the neutrino energy. 
The track like topology  gives a good estimation of the neutrino direction but suffer on a very large imprecision on the energy. 

In the shower-like events we have interactions of the $\nu_e$, $\nu_\tau$, interactions of neutral current: all these configurations generate a cascade which produces a large number of particles in a region of size of 10$\div$20 m; most of the energy of the neutrinos is dissipated in a small size in point-like interaction inside the neutrino telescope.
The neutrino cannot interact very far away from the detector, 
otherwise the light cannot arrive: 
the neutrino interactions of this kind of flavors can occur only inside the detector and there is a calorimetric measurement of the energy loss of the particles: the shower like topology gives a good estimation of the energy but suffer on a very large imprecision of the measurement of the incoming direction of neutrino because the neutrino track is very short with respect to the other dimensions of the detector.

\begin{figure}[!ht]
    \centering
    \includegraphics[width=0.7\textwidth]{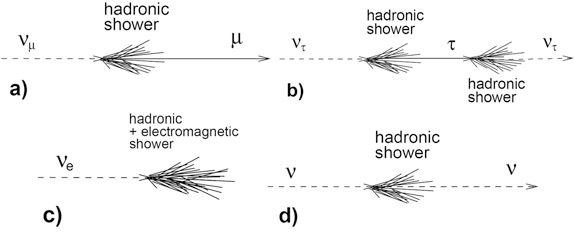}
    \setcaptionwidth{0.8\textwidth}
    \caption{\footnotesize{Some event signature topologies for different neutrino flavors and interactions: \textbf{a)} charged current interaction of a $\nu_\mu$ produces a muon and a hadronic shower; \textbf{b)} charged current interaction of a $\nu_\tau$ produces a $\tau$ that decays into a $\nu_\tau$, tracing the double bang event signature.  \textbf{c)} charged current interaction of $\nu_e$ produces both an EM and a hadronic shower; \textbf{d)} a neutral current interaction produces a hadronic shower. Particles and anti-particles cannot be distinguished in large volume neutrino detectors.}}
    \label{tr}
\end{figure}

A schematic views of $\nu_e$, $\nu_\mu$ and $\nu_\tau$ charged current events and of a neutral current events are shown in Fig. \ref{tr}.

\begin{figure}[!ht]
    \centering
    \includegraphics[width=0.5\textwidth]{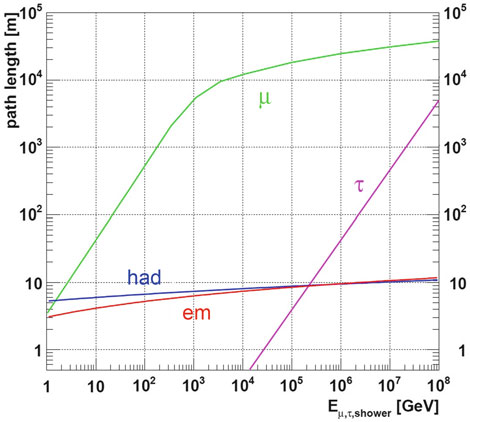}
    \setcaptionwidth{0.8\textwidth}
    \caption{\footnotesize{Path length of particles produced by neutrino interactions in water: muons ($\mu$), taus ($\tau$), electromagnetic (em) and hadronic (had) showers, versus their respective energy. The shower lengths are calculated using a shower profile parameterization.}}
    \label{pl}
\end{figure}

In Fig. \ref{pl} it is shown how long a muon can travel in water at a given energy (for instance, a 1 Tev muon can travel 5 km while hadronic and electromagnetic cascade scale lengths are less than 10 meters).


The effects of the medium (water or ice) on light propagation are absorption and
scattering of photons. These affect the reconstruction capabilities of the telescope.
Water is transparent only to a narrow range of wavelengths ($350\leq \lambda\leq 550$ nm).
The absorption length is the distance over which the light intensity has dropped to
$e^{-1}$ of the initial intensity $I_0$. Thus, the light intensity in a homogeneous medium reduces to a factor $I(x)=I_0^{−x/\lambda_{abs}}$ after traveling a distance $x$. For deep polar ice, the maximum value of $\lambda_{abs} ∼100$ m is assumed in the blue-UV region, while it is about 70 m for clear ocean waters. Absorption reduces the amplitude of the Cherenkov wavefront, i.e., the total amount of light arriving on photomultiplier tubes. Scattering changes the direction of the Cherenkov photons, and consequently delays their arrival time on the photomultiplier tubes; this degrades the measurement of the direction of the incoming neutrino. Direct photons are those arriving on a photomultiplier tube in the Cherenkov wavefront, without being scattered; the others are indirect photons. Seawater has a smaller value of the absorption length with respect to deep ice,
which is more transparent. The same instrumented volume of ice corresponds to a
larger effective volume with respect to seawater. On the other hand, the effective
scattering length for ice is smaller than water. This is a cause of a larger degradation of the angular resolution of detected neutrino-induced muons in ice with respect to water. By the time-space correlation of light collected by photomultiplier tubes put in the water we can reconstruct the direction and the estimate energy of the incoming neutrino and the total cost of the detector depends on how many modules are needed in order to find a track of neutrino in 1 $\mathrm{km}^3$ of water or ice. 

\begin{figure}[!ht]
    \centering
    \includegraphics[width=0.8\textwidth]{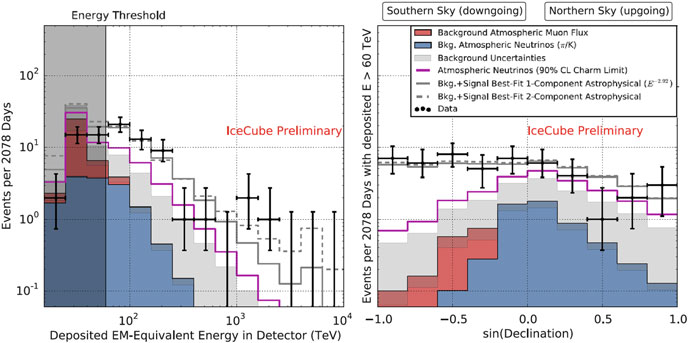}
    \setcaptionwidth{0.8\textwidth}
    \caption{\footnotesize{Deposited energies $E_{dep}$ (left panel) and arrival directions (right panel) of observed IceCube events (crosses), compared with predictions. The sample refers to 6 years of data. The hashed region shows uncertainties on the sum of all backgrounds, due to atmospheric muons and neutrinos. The contribution of an astrophysical ($\nu+\Bar{\nu}$) flux for $E_{dep} > 60$ TeV is also indicated. See Aartsen et al. (2017) \cite{2017arXiv171001191I} for details. Courtesy of the IceCube Collaboration.}}
    \label{back}
\end{figure}

The first observation of an excess of high-energy astrophysical neutrinos over the
expected background has been reported by IceCube and this sample is
continuously updated (\cite{2013ApJ...779..132A}) as shown in Fig. \ref{back}.

\begin{figure}[!ht]
    \centering
    \includegraphics[width=0.8\textwidth]{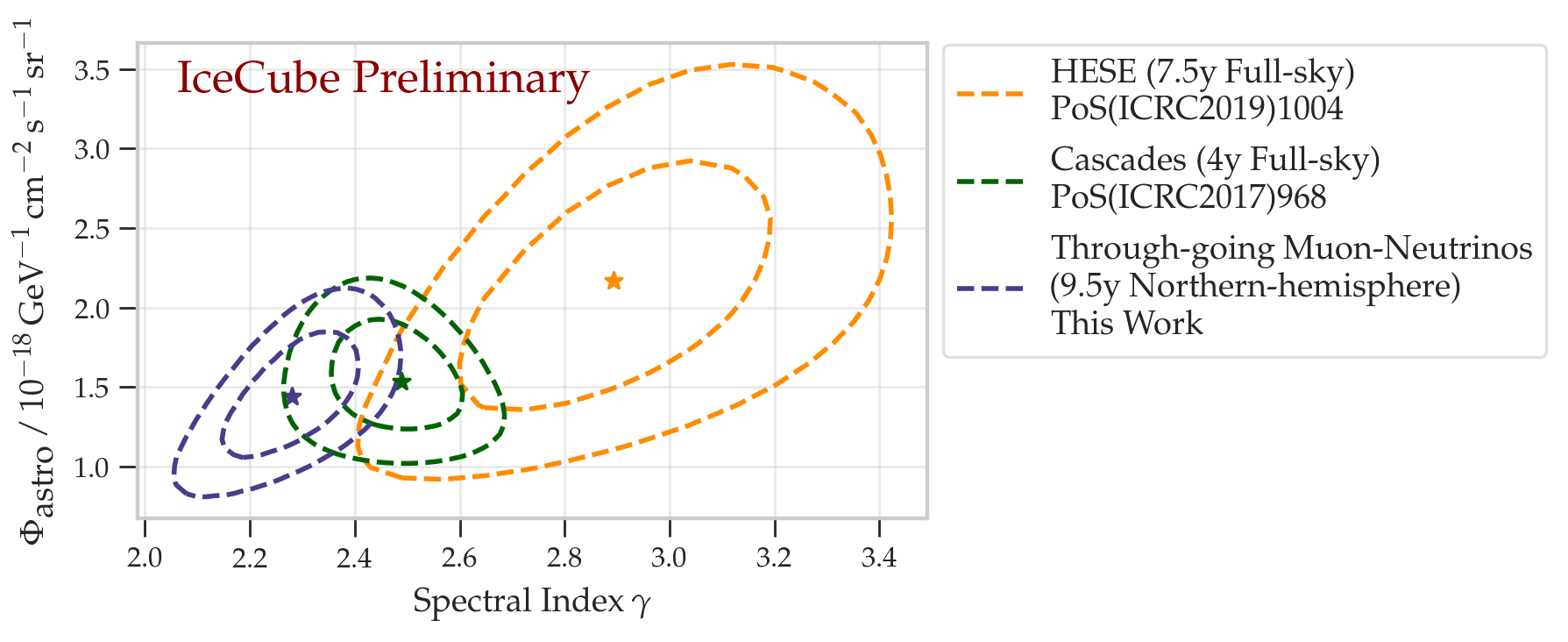}
    \setcaptionwidth{0.8\textwidth}
    \caption{\footnotesize{Comparison of the best-fit parameters for the single power-law model and their uncertainty contours: IceCube (HESE, Cascades and this work) and ANTARES (Tracks+Cascades, best-fit only).}}
    \label{pre}
\end{figure}

For extragalactic neutrinos, a hard energy spectrum $\phi_\nu(E)\propto E^{-\Gamma_\nu}$ with $\Gamma_\nu ∼ 2$, is motivated by models of cosmic ray production at sources within the framework of the Fermi acceleration mechanism. The IceCube spectrum measured through passing muons, has spectral index $\Gamma_\nu ∼ 2.2$, close to the expected value. These events originate from the Northern sky, where the presence of the Galactic plane is marginal. Most of them could likely be of extragalactic origin.
On the other hand, the sample of so-called High Energy Starting Events (HESE) presents a significantly softer spectrum, with $\Gamma_\nu ∼ 2.9$. The discrepancy between the best-fit for cosmic neutrinos from
HESE and from the passing muon sample presents an approximately $3\sigma$ tension if a single unbroken power law is assumed (\cite{2016ApJ...833....3A}). The possible origin of
this discrepancy (the presence of two different extragalactic components; a Galactic
plus an extragalactic component, etc.) is, at present, one intriguing research field in neutrino astrophysics. In Fig. \ref{pre} there is the comparison of the best-fit parameters for the single power-law model and their uncertainty contours.

\begin{figure}[!ht]
    \centering
    \includegraphics[height=0.45\textwidth]{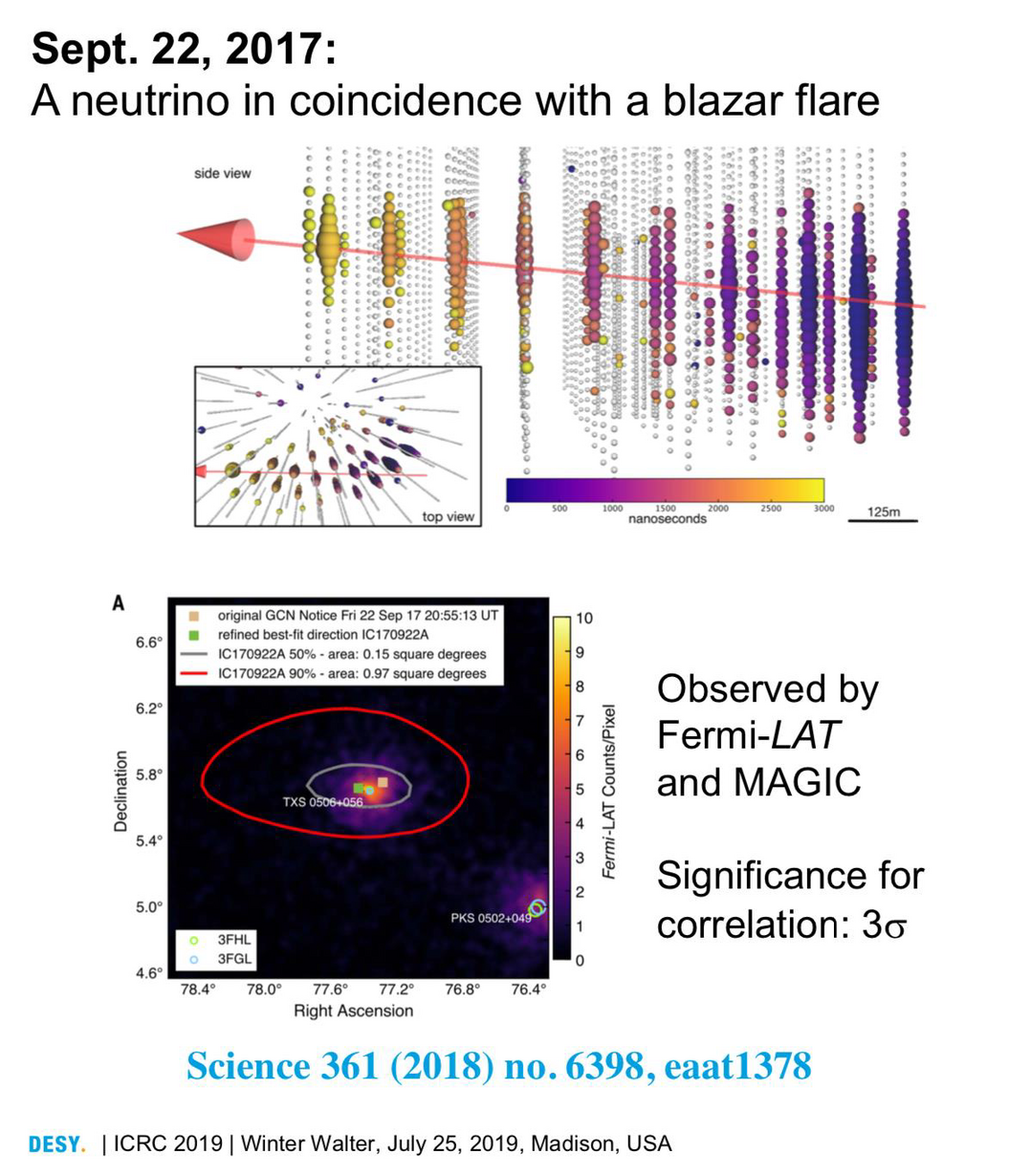}
    \includegraphics[height=0.45\textwidth]{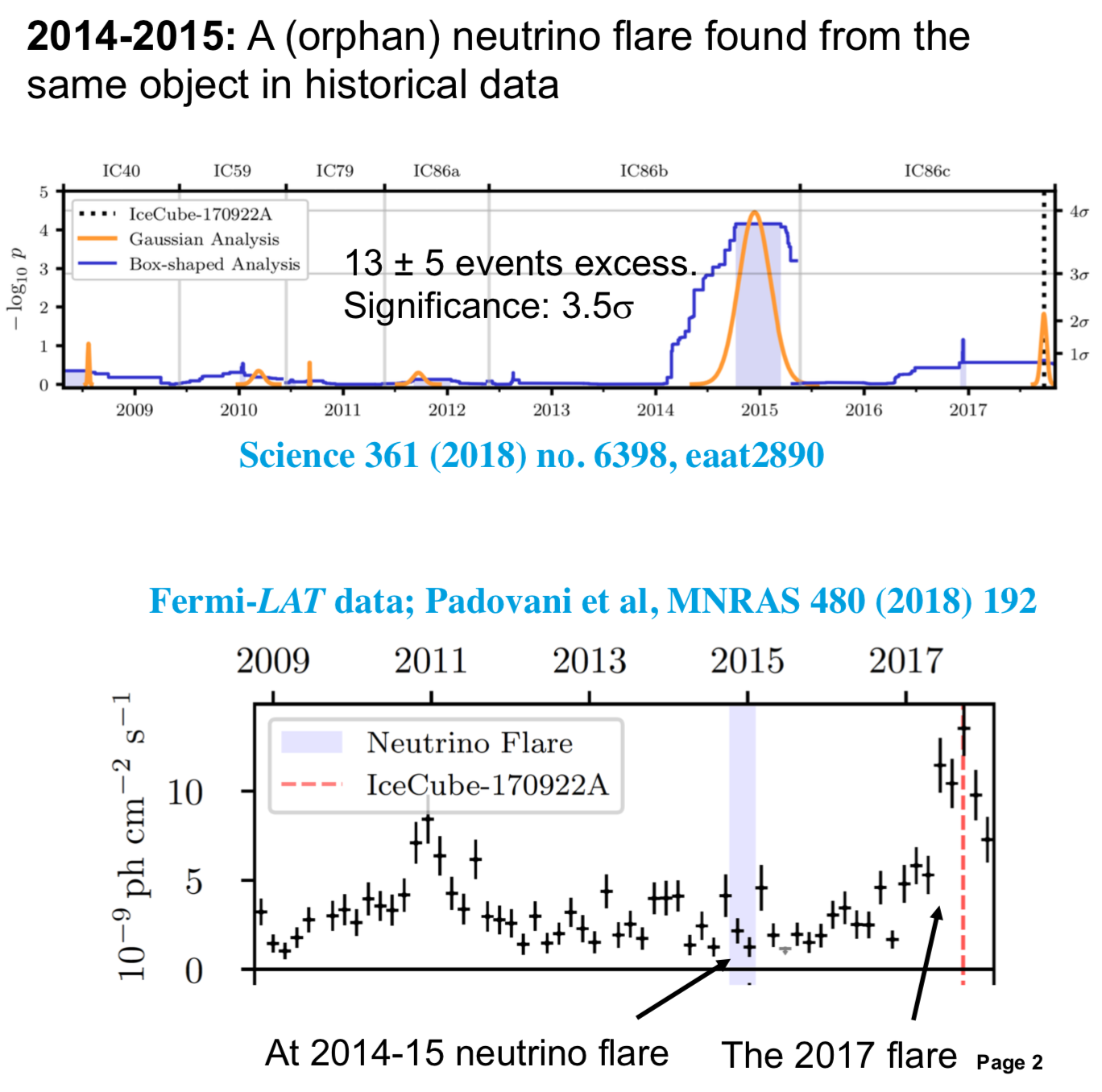}
    \setcaptionwidth{0.8\textwidth}
    \caption{\footnotesize{A neutrino coincidence with a blazar flare on September 22, 2017 \cite{2018Sci...361.1378I} and a neutrino flare found from the same object in historical data in 2014-15 \cite{doi:10.1126/science.aat2890,10.1093/mnras/sty1852}.}}
    \label{bl}
\end{figure}

On September 22, 2017, the IceCube Neutrino Observatory detected a high energy muon neutrino: IceCube-170922A that carried an energy of 290 TeV. The campaign indicated that this event came from
the direction of a known active galactic nucleus blazar: TXS $0506+056$, which was found to be in a flaring state of high $\gamma$-ray emission. It was subsequently observed at other wavelengths of light across the electromagnetic spectrum, including radio, infrared, optical, X-rays and $\gamma$-rays. $TXS \  0506+056$ is a BL Lac object, found at redshift $z=0.3365\pm0.0010$, and it was monitored by FERMILAT and observed by MAGIC. After the coincident event, a $\nu$-flare but without
associated $\gamma$-rays was found in archival IceCube data. A further analysis of archival IceCube data revealed that this blazar was emitting neutrinos before: within October 2014-March 2015 an excess of 13±5 events over background was found, see Fig. \ref{bl}. During this period, there was no significant EM flaring activity. This led to a not simple theoretical interpretation but the IceCube conclusion is for a compelling evidence of a high-energy neutrino from a blazar. The detection of both neutrinos and light from the same object was an early example of multi-messenger astronomy.

The observation of these exciting events was immediately followed by the community of  telescopes; for this communication there are devoted systems used as brokers to receive and transmit information in real time to all experiments so that the success of multi-messenger astronomy is also due to the use of this technique in order to have very prompt information of interesting events. The GCN (Gamma-ray Coordinates Network) distributes the locations of $\gamma$-ray bursts and other Transients (the Notices) detected by spacecraft (most in real-time while the burst is still bursting and others that are delayed due to telemetry down-link delays) and reports of follow-up observations (the Circulars) made by ground-based and space-based optical, radio, X-ray, TeV, and other particle observers. The Astrophysical Multi-messenger Observatory Network (AMON \cite{2013APh....45...56S}) is a program currently under development at The Pennsylvania State University, in collaboration with a growing list of U.S. and international observatories. AMON seeks to perform a real-time correlation analysis of the high-energy signals across all known astronomical messengers – photons, neutrinos, cosmic rays, and gravitational waves.

\begin{figure}[!ht]
    \centering
    \includegraphics[width=0.7\textwidth]{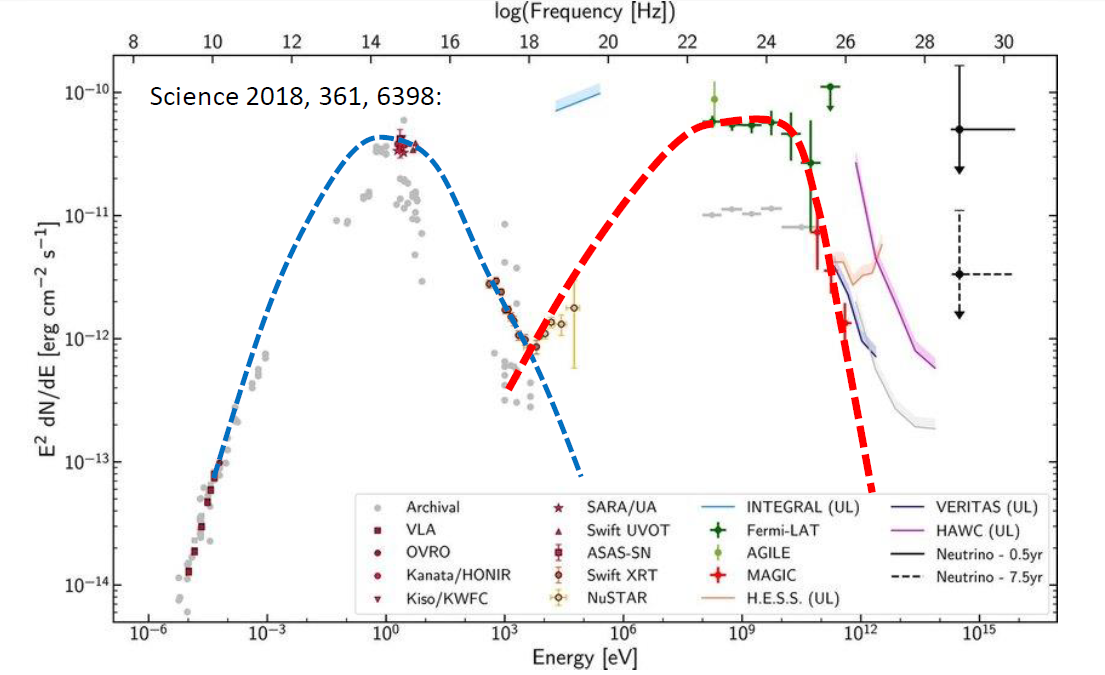}
    \setcaptionwidth{0.8\textwidth}
    \label{en}
    \caption{\footnotesize{Multi-messenger observations of a flaring blazar coincident with high-energy neutrino IceCube-170922A \cite{2018Sci...361.1378I}.}}
\end{figure}

\begin{figure}[!ht]
    \centering
    \includegraphics[height=0.3\textwidth]{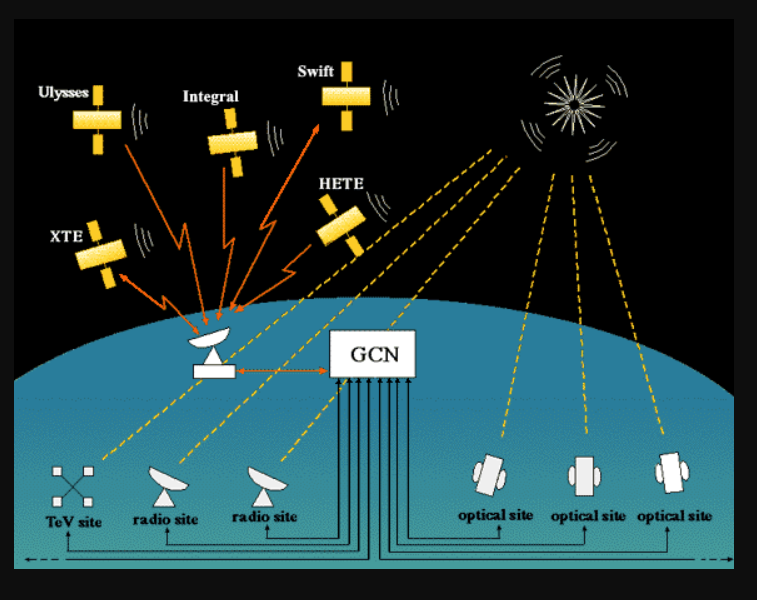}
    \includegraphics[height=0.3\textwidth]{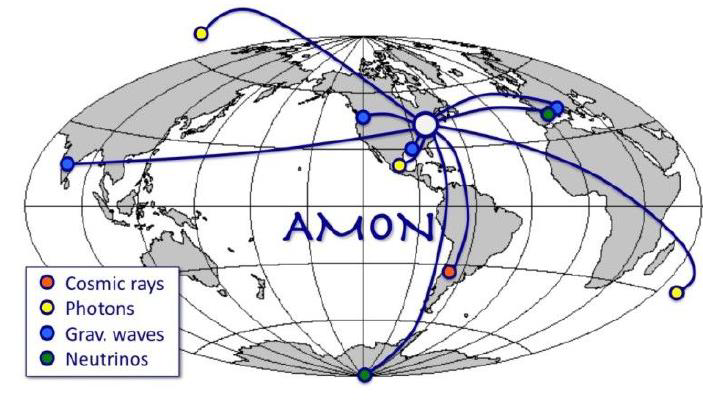}
    \setcaptionwidth{0.8\textwidth}
    \caption{\footnotesize{The GCN (Gamma-ray Coordinates Network) and the Astrophysical Multi-messenger Observatory Network (AMON).}}
    \label{amon}
\end{figure}

\begin{figure}[!ht]
    \centering
    \includegraphics[width=0.5\textwidth]{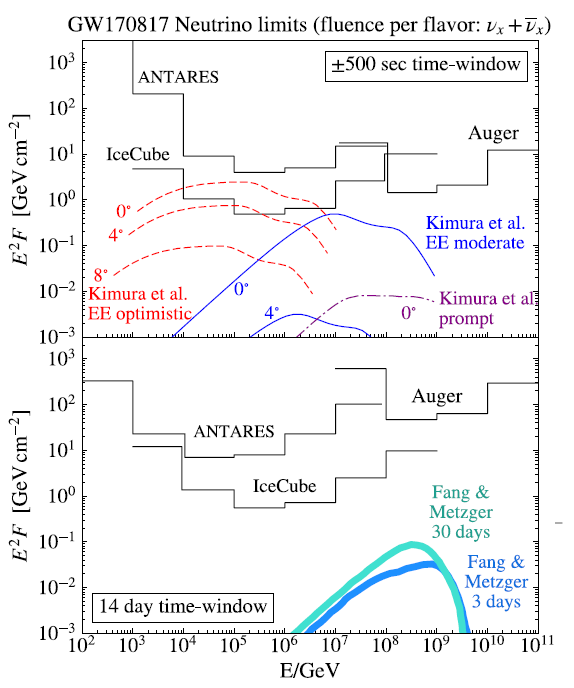}
    \includegraphics[width=0.5\textwidth]{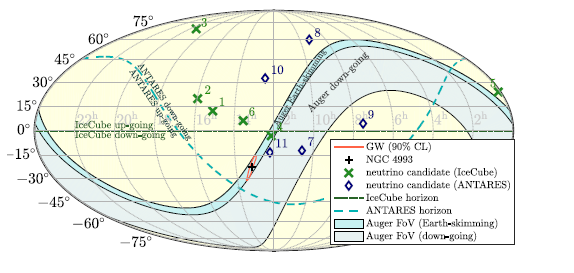}
    \setcaptionwidth{0.8\textwidth}
    \caption{\footnotesize{ANTARES, IceCube, Auger \& LVC, \cite{2017ApJ...850L..35A}.}}
    \label{obs}
\end{figure}

\subsection{Possible correlation between neutrinos and other events}

In addition to the correlation of neutrinos and EM radiation as the IceCube events mentioned before, there are also indications of tidal disruption events in which one star is totally destroyed during the passage close to a supermassive black hole: few events of this kind were observed with a mechanism in which a relativistic hadronic jet is formed in extreme cases, producing secondary neutrinos; the discussion among scientists is still in progress. There is also the observation of $\gamma$-rays and gravitational waves but no observation of neutrinos in these kinds of events because they are expected to be produced in a particular direction of the emission of the jet: merging neutron stars can emit a jet perpendicular to the plane of merging, far away to the axes in order to detect neutrinos. 

In the multi-messenger study of counterpart for the gravitational wave observed in coincidence with the neutron stars coalescence of 17 August 2017 (GW170817), about 80 detectors around the World have been involved. They include telescopes for infrared, optical and radio astronomy on the Earth ground, UV, X-ray and $\gamma$ ray telescopes on satellites, and neutrino observatories. In Fig. \ref{obs} we present the upper limits on neutrino emission connected with GW170817 obtained by IceCube, ANTARES and Pierre Auger.

\section{Conclusions} \label{conclusioni}

Multi-messenger astronomy is an increasing field in which methods to analyse data are improving. At the beginning of this century particle physics were done only using cosmic rays and after the 50s, with the construction of the first accelerators, astroparticle physics and particle physics were disentangled and they became different subjects.

From the 1950s, the study of the microcosm had an impressive growth, forced by the increasing energy of accelerators from the MeV to the TeV scale. To go far beyond the energy scale (10 TeV) reached by the LHC, efforts at the limit of human and financial possibilities are needed. The return to the use of cosmic accelerators will probably be a necessity.
From the particle physics point of view, the possibility of using cosmic beams to improve our understanding of Nature will depend upon either the detailed understanding of cosmic acceleration and on the development of methods for controlling systematic errors introduced by our lack of understanding of these processes.

Thus, the combined information arising from gravitational waves, from the measurements of $\gamma$-rays with high-resolution instruments, from high-statistics measurements of charged cosmic rays and from neutrino telescopes is mandatory to understand the nature of cosmic accelerators. For these purposes, multi-messenger observations are not just an advantage, but also a necessity.

\bibliographystyle{JHEP.bst}
\bibliography{bib}

\acknowledgments This work has been partially supported by Agencia Estatal de Investigaci\'{o}n (Spain) under 232 grant PID2019-106802GB-I00/AEI/10.13039/501100011033.

\end{document}